\documentclass{aa}  

\usepackage{natbib}
\usepackage{graphicx}
\usepackage{txfonts}
\usepackage[colorlinks=true,urlcolor=blue,linkcolor=blue,citecolor=blue]{hyperref}
\usepackage{listings}
\usepackage{amsmath}
\usepackage[normalem]{ulem}

\begin{document} 
  \title{Photometric redshift estimation via deep learning}
  \subtitle{Generalized and pre-classification-less, image based, fully probabilistic redshifts}
  \author{A.~D'Isanto\inst{1}
     \and
          K.L.~Polsterer\inst{1}
          }

  \institute{Astroinformatics, Heidelberg Institute for Theoretical Studies,
             Schloss-Wolfsbrunnenweg 35, 69118 Heidelberg, Germany\\
              \email{antonio.disanto@h-its.org},
              \email{kai.polsterer@h-its.org}
             }

  \date{Received } 

  \abstract
   {
   The need to analyze the available large synoptic multi-band surveys drives the development of new data-analysis methods.
   Photometric redshift estimation is one field of application where such new methods improved the results, substantially.
   Up to now, the vast majority of applied redshift estimation methods have utilized photometric features.
   }
   {
   We aim to develop a method to derive probabilistic photometric redshift directly from multi-band imaging data, rendering pre-classification of objects and feature extraction obsolete.
   }
   {
   A modified version of a deep convolutional network was combined with a mixture density network.
   The estimates are expressed as Gaussian mixture models representing the probability density functions (PDFs) in the redshift space.
   In addition to the traditional scores, the continuous ranked probability score (CRPS) and the probability integral transform (PIT) were applied as performance criteria.
   We have adopted a feature based random forest and a plain mixture density network to compare performances on experiments with data from SDSS (DR9).
   }
   {
   We show that the proposed method is able to predict redshift PDFs independently from the type of source, for example galaxies, quasars or stars.
   Thereby the prediction performance is better than both presented reference methods and is comparable to results from the literature.
   }
   {
   The presented method is extremely general and allows us to solve of any kind of probabilistic regression problems based on imaging data, for example estimating metallicity or star formation rate of galaxies.
   This kind of methodology is tremendously important for the next generation of surveys.
   }


\maketitle


\section{Introduction}

In recent years, the availability of large synoptic multi-band surveys increased the need of new and more efficient data analysis methods.
The astronomical community is currently experiencing a data deluge.
Machine learning and, in particular, deep-learning technologies are increasing in popularity and can deliver a solution to automate complex tasks on large data sets.
Similar trends can be observed in the business sector for companies like Google and Facebook.
In astronomy, machine learning techniques have been applied to many different uses \citep{2010ijmpd..19.1049b}. 
Redshift estimation is just one relevant field of application for these statistical methods.
Constant improvements in performances had been achieved by adopting and modifying machine learning approaches.
The need for precise redshift estimation is increasing due to its importance in cosmology \citep{doi:10.1111/j.1365-2966.2005.09526.x}.
For example, the Euclid mission \citep{2012spie.8442e..0tl} is highly dependent on accurate redshift values.
Unfortunately, measuring the redshift directly is a time consuming and expensive task as strong spectral features have to be clearly recognized.
Therefore, redshifts extracted via photometry based models provide a good alternative \citep{2016MNRAS.460.1371B}.
At the price of a lower accuracy compared to the spectroscopic measurements, photometric redshift estimates enable the processing of huge numbers of sources \citep{doi:10.1111/j.1365-2966.2011.19375.x}.
Moreover, by combining photometric and spectroscopic techniques, low signal-to-noise spectra of faint objects can be better calibrated and processed \citep{2001ApJS..135...41F}.

Photometric redshift estimation methods found in the literature can be divided in two categories:
template based spectral energy distribution (SED) fitting \citep[e.g.][]{2000A&A...363..476B, 0004-637X-690-2-1250} and statistical and/or machine learning algorithms \citep[e.g.][]{0004-637X-536-2-571, 2011mnras.418.2165l}.
In this work we will focus on the latter ones and in particular on the application of deep-learning models.
Most machine learning approaches demand a large knowledge base to train the model.
Once trained, such models allow us to process huge amounts of data and to automatically generate catalogs with millions of sources \citep[as done in][]{Brescia2014}.
Instead of generating point estimates only, extracting density distributions that grant access to the uncertainty in the prediction is gaining more focus in the community \citep{2013MNRAS.432.1483C}.
This scheme is increasingly important for the success of the Euclid mission, which depends on highly precise and affordable probabilistic photometric redshifts \citep{2017IAUS..325...73D}.

Due to the advent of faster and more specialized compute architectures as well as improvements in the design and optimization of very deep and large artificial neural networks, more complex tasks could be solved.
In recent years, the so called field of deep-learning was hyped together with big-data-analytics, both in the commercial sector as well as in science.
Astronomy has always been a data-intense science but the next generation of survey missions will generate a data-tsunami.
Projects such as the Large Synoptic Survey Telescope (LSST) as well as the Square Kilometre Array (SKA) are just some examples that demand processing and storage of data in the peta- and exabyte regime.
Deep-learning models could provide a possible solution to analyze those data-sets, even though those large networks are lacking the ability to be completely interpretable by humans.

In this work, the challenge of deriving redshift values from photometric data is addressed.
Besides template fitting approaches, several machine-learning models have been used in the past to deal with this problem \citep{2004PASP..116..345C, 2011mnras.418.2165l, 2013mnras.428..226p, Cavuoti20153100, 2016A&C....16...34H}.
Up to now, the estimation of photometric redshifts was mostly based on features that had been extracted in advance.
A Le-Net deep convolutional network (DCN) \citep{lecun-98} is able to automatically derive a set of features based on imaging data and thereby make the extraction of tailored features obsolete.
Most machine-learning based photometric redshift estimation approaches found in the literature just generate single value estimates.
Template fitting approaches typically deliver a probability density function (PDF) in the redshift space.
Therefore currently the machine learning based approaches are either modified or redesigned to deliver PDFs as well \citep{2016PASP..128j4502S}.
We have proposed a combined model of a mixture density network (MDN) \citep{Bishop94mixturedensity} and a DCN.
The MDN hereby replaces the fully-connected feed-forward part of the DCN to directly generate density distributions as output.
This enables the model to use images directly as input and thereby utilize more information in contrast to using a condensed and restricted feature-based input set.
The convolutional part of the DCN automatically extracts useful information from the images which are used as inputs for the MDN.
The conceptual idea of combining a DCN with a MDN together with details on the implementation have been presented to the computer science community in \citet{disanto2017esann} while this publication addresses the challenges of photometric redshift estimation on real world data by using the proposed method.
A broad range of network structures have been evaluated with respect to the performed experiments.
The layout of the best performing one is presented here.

The performance of the proposed image-based model is compared to two feature-based reference methods: a modified version of the widely used random forest (RF) \citep{Breiman:2001:RF:570181.570182} which is able to produce PDFs and a plain MDN.
The predicted photometric redshifts are expressed as PDFs using a Gaussian mixture model (GMM).
This allows the capture of the uncertainty of the prediction in a compact format.
Due to degeneracies in the physical problem of identifying a complex spectral energy distribution with a few broadband filters only, multi-modal results are expected.
A single Gaussian is not enough to represent the photometric redshift PDF.
When using PDFs instead of point estimates, proper statistical analysis tools must be used, taking into account the probabilistic representation.
The continuous ranked probability score (CRPS) \citep{2000WtFor..15..559H} is a proper score that is used in the field of weather forecasting and expresses how well a predicted PDF represents the true value.
In this work, the CRPS reflects how well the photometrically estimate PDF represents the spectroscopically measured redshift value.
The CRPS is calculated as the integral taken over the squared difference between the cumulative distribution function (CDF) and the Heaviside step function of the true redshift value.
In contrast to the likelihood, the CRPS is more focused on the location and not on the sharpness of the prediction.
We have adopted the probability integral transform (PIT) \citep{2005MWRv..133.1098G} to check the sharpness and calibration of the predicted PDFs.
By plotting a histogram of the CDF values at the true redshift over all predictions, overdispersion, underdispersion and biases can be visually detected.
We demonstrate that the CRPS and the PIT are proper tools with respect to the traditional scores used in astronomy.
A detailed description of how to calculate and evaluate the CRPS and the PIT are given in Appendix~\ref{sec:crps_and_pit}.

The experiments were performed using data from the Sloan Digital Sky Survey (SDSS-DR9) \citep{2012ApJS..203...21A}.
A combination of the SDSS-DR7 Quasar Catalog \citep{2010AJ....139.2360S} and the SDSS-DR9 Quasar Catalog \citep{2012A&A...548A..66P} as well as two subsamples of $200,000$ galaxies and $200,000$ stars from SDSS-DR9 are used in Sect.~\ref{sec:experiments}.

\vfill
{\bf outline:}
In Sect.~\ref{sec:models} the machine learning models used in this work are described.
The layout of the experiments and the used data are described in Sect.~\ref{sec:experiments}.
Next, in Sect.~\ref{sec:results} the results of the experiments are presented and analyzed.
Finally, in Sect.~\ref{sec:conclusions} a summary of the whole work is given.
In the appendix, an introduction on CRPS and PIT is given, alongside with the discussion of the applied loss function.
Furthermore, we motivate the choice of the number of components in the GMM used to describe the predicted PDFs.
Finally, the SQL queries used to download the training and test data as well as references to the developed code are listed.


\section{Models}\label{sec:models}
In this section the different models used for the experiments are described.
The RF is used as a basic reference method while the MDN and the DCN are used to compare the difference between feature based and image based approaches.

\subsection{Random forest}\label{sec:rf}
The RF is a widely used supervized model in astronomy to solve regression and classification problems.
By partitioning the high dimensional features space, predictions can be efficiently generated.
Bagging as well as bootstrapping make those ensembles of decision trees statistically very robust.
Therefore RF is often used as a reference method to compare performances of new approaches.
The RF is intended to be used on a set of input features which could be plain magnitudes or colors in the case of photometric redshift estimation \citep{2010ApJ...712..511C}.
In its original design, the RF does not generate density distributions.
To produce PDFs, the results of every decision tree in the forest are collected and a Gaussian mixture model (GMM) is fitted.
The PDF is then presented by a mixture of $n$ components:

\vfill
\begin{equation}\label{eq:gmm}
p(x) = \sum_{j=1}^{n}\omega_{j}\mathcal{N}(x|\mu_{j}, \sigma_{j}),
\end{equation}
\vfill

\noindent where $\mathcal{N}(x|\mu_{j}, \sigma_{j})$ is a normal distribution with a given mean $\mu_{j}$ and standard deviation $\mu_{j}$ at a given value $x$.
Each component is weighted by $\omega_{j}$ and all weights sum to one.
To calculate the CRPS for a GMM, the equations by \citet{2006QJRMS.132.2925G} can be used.
For the proposed experiments a model composed of $256$ trees is used.
As input the five $ugriz$ magnitudes and the pairwise color combinations are used, obtaining a feature vector of $15$ dimensions per data item.

\subsection{Mixture density network}\label{sec:mdn}

\begin{figure*}
\centering
  \includegraphics[trim=0 0 0 0, clip, width=0.85\textwidth]{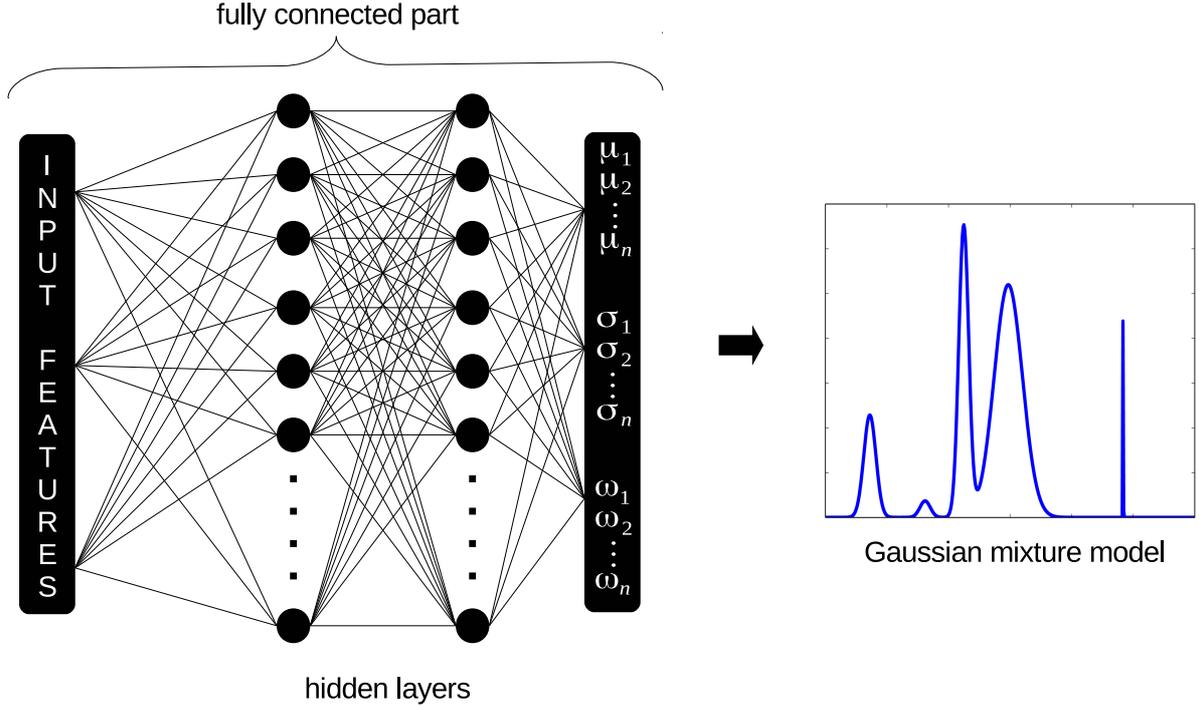}
    \caption{Architecture of the mixture density network.
    Next to the input layer, two hidden and fully interconnected layers are depicted. As output a vector for each parameter of the GMM is predicted ($\mathbf{\mu}$, $\mathbf{\sigma}$, $\mathbf{\omega}$).
    Based on this compact description, a density distribution can be generated.}\label{Fig:mdn}
\end{figure*}

An MDN \citep{Bishop94mixturedensity} is a modified version of the widely known multilayer perceptron (MLP) \citep{rosenblatt1962principles}, producing distributions instead of point estimates.
The MLP is a supervized feed-forward neural network, whose main calculation unit is called neuron.
The neurons are arranged in layers, an input layer, one or more hidden layers, and an output layer.
Several hyperparameters characterize the architecture of an MLP.
The activation function is a non linear function applied to the data as they pass through the neurons.
Commonly a sigmoidal function or hyperbolic tangent (\emph{tanh}) are utilized.
In recent years, the MLP has been widely used in astronomy, for example to estimate redshifts based on photometric features \citep{Brescia2013}.

The MDN interprets the output vector $o$ of the network as the means $\mathbf{\mu}$, standard deviations $\mathbf{\sigma}$ and weights $\mathbf{\omega}$ of a mixture model, using $n$ Gaussians as basis functions:
\begin{equation}\label{eq:mu_sigma_omega}
\begin{split}
\mu_{j} &= o_{j}^{\mu},\\
\sigma_{j} &= \exp(o_{j}^{\sigma}),\\
\omega_{j} &= \frac{\exp(o_{j}^{\omega})}{\sum_{i=1}^{n}\exp(o_{i}^{\omega})} \\
 &\textrm{with } j \in 1...n \textrm{ and } o = \{o^\mu,o^\sigma,o^\omega\}.
\end{split}
\end{equation}
Commonly, MDNs are trained using the log-likelihood as loss function.
In this application the focus is more on the distribution and shape than on the location of the prediction; hence the CRPS is adopted as loss function.
A detailed analysis of the performances of both loss functions is provided in Appendix~\ref{sec:CRPSvslL}.
The CRPS increases linearly with the distance from the reference value while the log-likelihood increases with the square of the distance.
Like the RF, the MDN is a feature-based model and therefore can use exactly the same input features: plain magnitudes and colors.
A generalized architecture of an MDN is shown in Fig.~\ref{Fig:mdn}.

\subsection{Deep convolutional mixture density network}\label{sec:dcmdn}
A DCN is a feed-forward neural network in which several convolutional and sub-sampling layers are coupled with a fully-connected part.
In some sense, it is a specialization of the fully interconnected MLP model.
By locally limiting the used interconnections in the first part, a spatial sensitivity is realized.
Thereby the dimensionality of the inputs is reduced step-wise in every layer.
The second part makes use of the so-called feature maps in a flattened representation, using them as input for an ordinary MLP.

This kind of network architecture finds wide application in the fields of image, video and speech recognition due to its capability of performing some kind of dimensionality reduction and automatic feature extraction.
A DCN model was chosen because of its ability to deal directly with images, without the need of pre-processing and an explicit features extraction process.
The network is trained to capture the important aspects of the input data.
By optimizing the condensed representation of the input data in the feature maps, the performance of the fully connected part is improved.
Every input data item is a tensor of multi-band images.
In the experiments the five $ugriz$ filters from SDSS as well as the pixel-wise differences were used.
Those image gradients can be compared to the colors in feature based approaches that minimize the effects of object intrinsic variations in luminosity.
In the convolutional layers, the input images are spatially convolved with proper filters of fixed dimensions.
Hereby 'proper' denotes a size that corresponds to the expected spatial extension of distinctive characteristics (filter of the dimension $3 \times 3$ and $2 \times 2$ have been selected for the application at hand).
The filters constitute the receptive field of the model.

\begin{figure*}
\centering
  \includegraphics[trim=0 0 0 0, clip, width=1.00\textwidth]{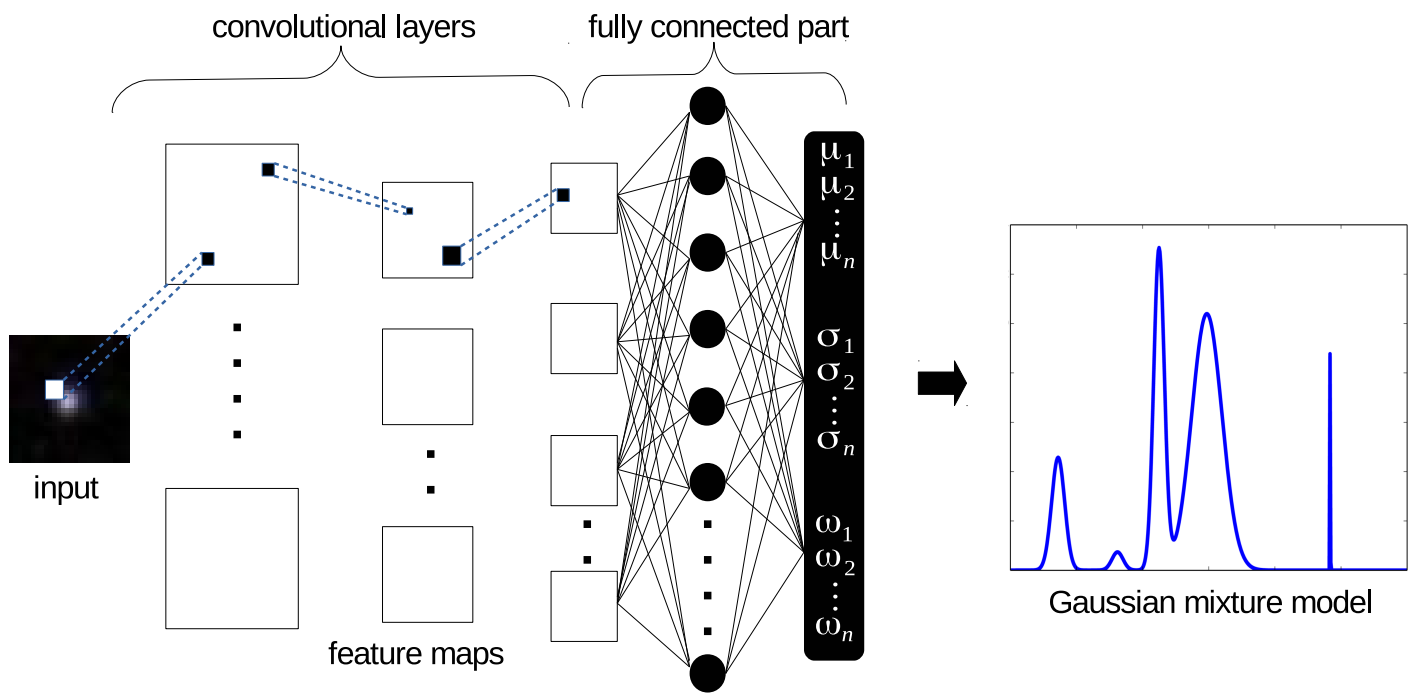}
    \caption{Architecture of the deep convolutional mixture density network.
    This architecture directly uses image matrices as input.
    In this figure, three convolutional layers are drawn, showing the local connection between the different layers.
    In the fully connected part, a MDN with one hidden layer is used and a vector for each parameter of the GMM is given as result ($\mathbf{\mu}$, $\mathbf{\sigma}$, $\mathbf{\omega}$).
    A sample of a predicted PDF is shown exemplarily.}\label{Fig:dcmdn}
\end{figure*}

After the convolution with the filters, a non-linear function is applied.
The \emph{tanh} was used in this work as expressed by the following relation:
\begin{equation}
H_{k} = tanh(W_{k}*X + b_{k}).
\end{equation}
\noindent In general, for every filter $k$ the hidden activation $H_{k}$ is calculated by using the filter weight matrix $W_{k}$ and the filter bias $b_{k}$. 
The outputs of the previous layer $H$ are used as inputs $X$ for the succeeding convolutional layer.
The first layer is directly fed with the imaging data that should be processed.
This filtering system with local connection make the architecture sensitive to spatial input patterns.
The filters are tiled through the entire visual field, with a fixed stride, generating a new representation of the input, that is, the feature maps.
A useful tool to pursue the down sampling is to apply max-pooling between the convolutional layers.
Those pooling filter typically select only the maximum value in confined region, typically of dimension $2\times2$ or $3\times3$.
Convolutional and pooling layers are alternated until the desired complexity in terms of feature maps is obtained.
In addition, those layers are alternated with \emph{rectified linear unit} (\emph{ReLu}) layers, in which the non-linear activation function is substituted by a non saturating function:
\begin{equation}
H_{k} = max(0, W_{k}*X + b_{k}).
\end{equation}
This function has many advantages; in particular it increases the efficiency of the gradient descent process, avoiding vanishing gradient problems.
Furthermore, using only comparison, addition and multiplication operations it is computationally more efficient.
The choice of the activation functions, namely the \emph{tanh} and the \emph{rectified linear unit}, influences the convergence and performance of the neural network.
Therefore, the activation function has to be considered as a free parameter when designing the network architecture.
In our case, switching to the \emph{rectified linear unit} improved the performance of the predictions notably.
Many different combinations have been tested, choosing the best performing one.
The feature maps constitute the input for the fully connected part, which has in general the same behavior as a MLP.
In the proposed model, we substitute the usual fully connected part with a MDN in order to generate PDFs instead of single point estimates.
For this reason, this combined architecture is denoted by us as deep convolutional mixture density network (DCMDN).
The structure of the DCMDN is sketched in Fig~\ref{Fig:dcmdn}.
Furthermore, as for the MDN, the CRPS is used as loss function.

Several hyperparameters influence the layout of the network architecture as well as the training phase.
Multiple combinations have been tested extensively.
Due to the immense amount of possible parameter combinations the currently used solution was obtained by clever manual engineering.
The most influencing parameters are listed below:

\begin{itemize}
  \item{{\it global architecture}: This includes the number and types of layers in the local-connected part and the number of hidden layers and neurons characterizing the fully-connected part.}
  \item{{\it activation function}: Defines the non linear function to process the input values of a neuron.}
  \item{{\it number of filters}: Influences the number of generated feature maps and therefore the amount of extracted features.}
  \item{{\it filter shape}: Characterizes the dimensions of the filters used; it can vary from layer to layer.}
  \item{{\it max-pooling shape}: As for the filters, it specifies the dimension of the area to which the max-pooling is applied.}
  \item{{\it learning rate}: Influences the step-size during the gradient descent optimization. This value can decrease with the number of trained epochs.}
  \item{{\it number of epochs}: Defines how often the whole training data set is used to optimize the weights and biases.}
  \item{{\it batch size}: As a stochastic gradient descent optimization strategy was chosen, this number defines the amount of training patterns to be used in one training step.}
\end{itemize}

The presented model exhibits many advantages for the application to photometric redshift estimation tasks.
It can natively handle images as input and extract feature maps in a fully-automatized way.
The DCMDN does not need any sort of pre-processing and pre-classification on the input data and as shown in the experiments section, the performances are far better with respect to the reference models.
The reason for an improvement with respect to the estimation performance is the better and extensive use of the available information content of the images.
Besides automatically extracting features, their importance with respect to the redshift estimation task is automatically determined too.
Feature based approaches make use of the condensed information that is provided through previously extracted features only.
Those features are good for a broad set of applications but not optimized for machine learning methods applied to very specific tasks.

Depending on the size of the network architecture, an extremely high number of weights and biases has to be optimized.
This allows the network to adopt to not significant correlations in the data and therefore overfit.
While good performances are achieved on the training data, the application to another data-sets would exhibit poor results.
To limit the effect of overfitting, the dropout technique can be applied; randomly setting a given percentage of the weights in both parts of the network to zero.
As deep-learning methods highly benefit from a huge amount of training objects, data augmentation is a common technique.
By simply rotating the images by $90^\circ$, $180^\circ$ and $270^\circ$ and flipping the images, the size of the training data-set can be increased by a factor of four and eight, respectively.
This reduces the negative effects of background sources and artifacts on the prediction performance.
Moreover, the early stopping technique can be applied to limit the chance of overfitting.
As soon as the performance that is evaluated on a separate validation set starts to degrade while the training error is still improving, the training is stopped even before reaching the anticipated number of epochs.
The DCMDN is based on the LeNet-5 architecture \citep{lecun-98} and realized in Python, making use of the Theano library \citep{2016arXiv160502688short}.

The architecture implemented by us is meant to run on graphics processing units (GPUs) as the training on simple central processing units (CPUs) is by far too time consuming.
In our experiments a speedup of factor of $\approx40 \times$ between an eight-core CPU (i7 4710MQ $2.50$\,GHz$\times8$) and the GPU hardware was observed.
During the experiments a cluster equipped with Nvidia Titan X was intensively used, allowing us to evaluate a larger combination of network architectures and hyperparameters.


\section{Experiments}\label{sec:experiments}
In the following sections the experiments performed with the presented models are described.
Those experiments are intended to compare the probabilistic redshift prediction performances of different models on different data-sets.
The data-sets used for training the models as well as evaluating the performances are described in the following.

\subsection{Data}\label{sec:data}
\begin{figure}
\centering
  \includegraphics[width=0.90\columnwidth]{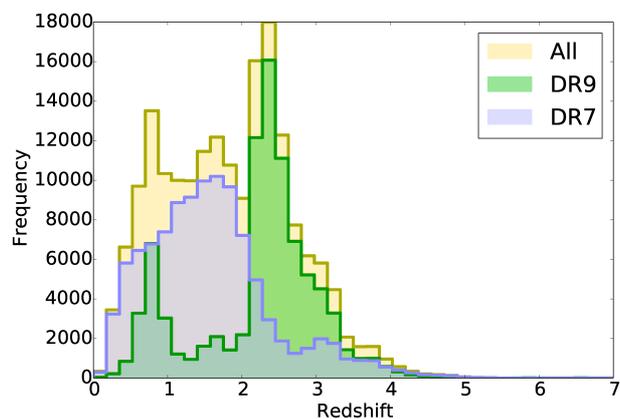}
    \caption{Redshift distribution of the quasar data from DR7, DR9 and the combined catalog, respectively.
    }
    \label{Fig:redshift_histo}
\end{figure}

All data-sets have been compiled using data from SDSS.
To generate a set of objects that cover the whole range of redshifts, separate data-sets for quasars and galaxies have been created.
The SDSS Quasar Catalog Seventh Data Release \citep{2010AJ....139.2360S}, containing $105,783$ spectroscopically confirmed quasars and the SDSS Quasar Catalog Ninth Data Release \citep{2012A&A...548A..66P}, containing $87,822$ spectroscopically confirmed quasars, are used as basis for the quasar data-set.
The two catalogs had to be combined because the former contains confirmed sources from SDSS II only, while the latter is composed of $91\%$ of new objects observed in SDSS III-BOSS \citep{2013AJ....145...10D}.
In this way a catalog with a much better coverage of the redshift space has been composed (see Fig.~\ref{Fig:redshift_histo}).
Furthermore, two samples composed of $200,000$ randomly chosen galaxies and $200,000$ randomly picked stars from DR9 (\cite{2012ApJS..203...21A}) have been selected (queries are stated in Appendix~\ref{sec:queries}).
The objects that have been classified by the spectroscopic pipeline as stars are assigned a redshift of zero.
These objects are mandatory to crosscheck the performance on objects that have not been pre-classified and therefore might be contaminated with stellar sources.

For every catalog the corresponding images from SDSS DR9 in the five $ugriz$ filters have been downloaded.
The images have been downloaded using the Hierarchical Progressive Surveys (HiPS) data partitioning format \citep{2015A&A...578A.114F}, in a \mbox{$28 \times 28$\,square-pixels} format, and subsequently pairwise subtracted, in order to obtain color images corresponding to the colors used in the feature-based experiments.
As described above, those pixel wise gradients minimize the effects of object intrinsic variations in luminosity.

\subsection{Setup and general configuration}\label{sec:design}
Three different categories of experiments have been performed, each using the three models of Sect.~\ref{sec:models} to compare the performances.
Before stating the details of the single experiments, a brief description of the general parameters adopted for the models is given.
The experiments with the RF and the MDN are feature-based, while the DCMDN is trained on the image representations of the data items.
As input features for the two reference models, magnitudes in the $ugriz$ filter are used together with all pair-wise color combinations.
For the RF models a fixed structure with 256 decision trees was chosen.
Bootstrapping was used for creating the individual trees.
Neither the depth nor the number of used features have been limited.
The MDN architectures use 15 inputs neurons (corresponding to the 15 input features) followed by two hidden layers containing 50 and 100 neurons, respectively.
15 output neurons are used to characterize the parameters of a GMM with $n=5$.
Photometrically estimated redshift distributions are of complex and multimodal nature \citep{KuglerGP16}.
The choice of 5 Gaussian components is based on experiments where the Bayesian information criterion (BIC) was used as a metric.
Depending on the redshift region calculated for, the BIC is indicating values between one and five components.
The results presented in Appendix~\ref{sec:BIC} indicate that five components are, on average, a good choice.
For the DCMDN, many different architectures have been tested, comprising less deep and compact convolutional parts.
The architecture that gave the best performances was finally created via a clever manual and empirical engineering.
The learning rate influences the size of the steps when applying gradient descent during the training with the backpropagation algorithm.
For the experiments presented in this work a \emph{tanh} was chosen with the learning rate degrading over the number of epochs.
To prevent overfitting, a common technique is to stop training as soon as the training and evaluation error starts to diverge (early stopping).
When training both different network architectures an early stopping rule was applied, varying the learning rate from a maximum of $0.01$ to a minimum of $0.001$ and changing the mini-batch size from $1,000$ to $500$.
In the DCMDN, dropout with a ratio of $60\%$ is applied to limit overfitting.

The evaluation of the performances of the models in the different experiments is done by using commonly used scores.
The root mean square error (RMSE), the median absolute deviation (MAD), the bias and the $\sigma^{2}$, are calculated based on the mean of the predicted redshift probability distributions and on the first (most significant) mode.
In addition, a normalized version is calculated by weighting the individual prediction errors with the true redshifts before calculating the final scores.
Moreover, in the experiments labeled as 'selected', the objects that show a complex and/or multi-modal behavior based on the predicted PDFs are excluded from the score calculation.
Those complex objects do not allow a single value as prediction result and therefore do not support a meaningful evaluation through the standard scores.
The results without the objects showing ambiguous predictions are presented in addition to the scores obtained on all objects with the previously mentioned performance measures.
To be more precise, only objects that fulfill
\begin{equation}\label{eq:constraints}
\begin{split}
|\mu_{1} - \mu_{2}| &< (\sigma_{1} + \sigma_{2})
\\ &\textrm{ or } \\
\omega_{1} &> 0.9545
\end{split}
,\end{equation}
\noindent have been selected.
Here $\mu_{1}$, $\mu_{2}$, $\sigma_{1}$ and $\sigma_{2}$ denote the means and the variances of the first two most significant modes of the predicted PDFs.
The modes are chosen based on their weights $\omega_j$, with $\omega_1$ being the weight of the strongest mode.
This selection criterion just picks objects where the first two modes are close to each other with respect to their widths or those objects with an extremely dominant first mode.
The value of $\omega_{1}$ was chosen a priori to ensure that $2\sigma$ of the joined distribution are represented by the dominant peak.

As PDFs need proper tools and scores for evaluation, the CRPS and PIT for every experiment are reported as well.
Those measures are much better able to report the performance of the estimations with respect to the spectroscopic redshifts.
In fact, they are capable of taking the location and the shape of the density distribution into account; important characteristics which the commonly used scores can not capture.
Typically a $k$-fold cross-validation is performed to prevent overfitting when training and evaluating machine learning models.
Due to the extreme computational demands when training the DCMDN, only a simple hold out strategy was used to evaluate the performance.
This is reasonable, as the same training, test and evaluation data-sets are used for all models.
The reported performances therefore allow a fair and qualitative comparison between the individual models, even though the presented absolute performances might slightly vary depending on the used training and testing data.
To be able to present an absolute performance which can be quantitatively compared to those in other publications, adequate reference data-sets have to be defined and published.
The architectures used for the DCMDN are stated in Tables \ref{tab:dcmdn_architecture_28x28} and \ref{tab:dcmdn_architecture_16x16}.
Two different input sizes for images have been used, depending on the experiment.
Larger images are used for galaxies while quasars and the mixed experiments use smaller images.
For this reason two different architectures are presented.

\begin{table}
  \centering
  \begin{tabular}{ c  l  c  c  c }
  \hline
  \# & Type & Size & \# Maps & Activation \\
  \hline
  1 & input & $15\times28\times28$ & / & / \\
  \hline 
  2 & convolutional & $3\times3$ & 256 & \emph{tanh} \\
  \hline
  3 & pooling & $2\times2$ & 256 & \emph{tanh} \\
  \hline
  4 & convolutional & $2\times2$ & 512 & \emph{tanh} \\
  \hline
  5 & pooling & $2\times2$ & 512 & \emph{tanh} \\
  \hline
  6 & convolutional & $3\times3$ & 512 & \emph{ReLu} \\
  \hline
  7 & convolutional & $2\times2$ & $1,024$ & \emph{ReLu} \\
  \hline
  8 & MDN & 500 & / &  \emph{tanh} \\
  \hline
  9 & MDN & 100 & / & \emph{tanh} \\
  \hline
  10 & output & 15 & / & Eq. \ref{eq:mu_sigma_omega} \\
  \hline
  \end{tabular}
  \caption{Layout of the DCMDN architecture for the experiments with the galaxies catalog. Stacks of 15 input images of the size \mbox{$28\times28$\,square-pixels} are used.}
\label{tab:dcmdn_architecture_28x28}
\end{table} 

\begin{table}
  \centering
  \begin{tabular}{ c  l  c  c  c }
  \hline
  \# & Type & Size & \# Maps & Activation \\
  \hline
  1 & input & $15\times16\times16$ & / & / \\
  \hline 
  2 & convolutional & $3\times3$ & 256 & \emph{tanh} \\
  \hline
  3 & pooling & $2\times2$ & 256 & \emph{tanh} \\
  \hline
  4 & convolutional & $2\times2$ & 512 & \emph{tanh} \\
  \hline
  5 & pooling & $2\times2$ & 512 & \emph{tanh} \\
  \hline
  6 & convolutional & $2\times2$ & $1,024$ & \emph{ReLu} \\
  \hline
  7 & MDN & 500 & / &  \emph{tanh} \\
  \hline
  8 & MDN & 100 & / & \emph{tanh} \\
  \hline
  9 & output & 15 & / & Eq. \ref{eq:mu_sigma_omega} \\
  \hline
  \end{tabular}
  \caption{Layout of the DCMDN architecture for the experiments with the quasar and the mixed catalog. Stacks of 15 input images of the size \mbox{$16\times16$\,square-pixelss} are used.}
\label{tab:dcmdn_architecture_16x16}
\end{table} 

\subsubsection*{Experiment 1 - Galaxies}
In the first experiment we perform the prediction of redshift PDFs on galaxies only.
The $200,000$ patterns contained in the galaxies catalog are split in a training and test-set, both containing $100,000$ objects each.
The images for all experiments have been cut-out with a size of \mbox{$28\times28$\,square-pixels}.
Only the galaxies experiment kept the original size.
As galaxies are extended objects this enables a better use of the available information.
Together with the five \emph{ugriz} images, all color combinations are built and hence a $15\times28\times28$ dimensional tensor is retrieved as object representation to be used as input for the DCMDN.
The architecture of the DCMDN that was used for this experiment is specified in Table \ref{tab:dcmdn_architecture_28x28}.
This first experiment is intended to test and compare the performances of the three models on objects in the low redshift range, taking into account the spatial extension.
As most objects of the galaxies sample are in a redshift range of $z\in[0..1]$, this provides a good testbed for the nearby Universe.
Experiment 1 is just based on galaxies and therefore a strong bias is introduced in the training phase.
The derived models are just producing usable redshift estimations when applied to images of galaxies.
Such a model is limited to objects with a correct and proper pre-classification and selection.

\subsubsection*{Experiment 2 - Quasars}
In the second experiment PDFs are estimated for quasars.
The quasar catalog is composed of $185,000$ objects; the DR7 and DR9 catalogs are combined selecting all the objects and removing double ones.
The quasar experiments have been performed using $100,000$ objects in the training-set and $85,000$ in the test-set.
This makes the size of the data used for training comparable to the previous experiment.
In this experiment the size of the input images for the DCMDN is reduced to \mbox{$16\times16$\,square-pixels} in order to save computational resources and speed up the training.
As quasars are more compact sources, a smaller cut-out should be sufficient to capture the details of the spatial distribution and still include information from the hosting galaxy.
The same color combinations used in the first experiment were created and a $15\times 16 \times 16$ tensor is used as input for the DCMDN.
The architecture of the DCMDN that was used for this experiment is specified in Table \ref{tab:dcmdn_architecture_16x16}.
The quasar experiment tests the performance for less extended objects that cover a wider range of redshift $z\in[0..6]$.
Similarly to the first experiment, the models of the second experiment were heavily dependent on a correct pre-classification of objects.

\subsubsection*{Experiment 3 - Mixed}
Finally, in the third experiment a mixed catalog was used.
By combining quasars, galaxies and stars we are able to test and evaluate the performances of the three models independently from the nature of the sources.
The step of pre-classifying objects is hereby made obsolete.
The stars that have been added can be considered as contamination; as faint cool stars can be easily confused with quasars.
This makes the use-case of photometric redshift estimation more realistic to the challenges of processing yet unseen objects with an uncertain classification.
To be able to derive a proper PDF for all objects, stars have been assigned a redshift of $z=0$.
As stated above, the whole catalog is composed of $585,000$ objects.
In this experiment, $300,000$ patterns were used for training and $285,000$ for testing and the dimension of the input images is reduced to \mbox{$16\times16$\,square-pixels}.
The DCMDN has therefore the same architecture as used in the previous experiment (see Table \ref{tab:dcmdn_architecture_16x16}).
This experiment is intended to evaluate the performances of the models in a realistic use-case.
Hence the results of this experiments are the most notable, as no biases through a pre-classification phase are introduced.
Such an experiment should be part of every publication, introducing a new data-driven method for photometric redshift estimation.

\section{Results}\label{sec:results}
The experiments of Sect.~\ref{sec:experiments} have been performed on a GPU cluster.
The detailed results are presented in the following.

\subsection{Experiment 1 - Galaxies}

\begin{figure*}
  {\footnotesize \hspace{3.7em} Mean, all $100,000$ objects \hspace{7.1em} Mean, all $100,000$ objects \hspace{7.1em} Mean, all $100,000$ objects \hfill\,}
  \includegraphics[width=\textwidth]{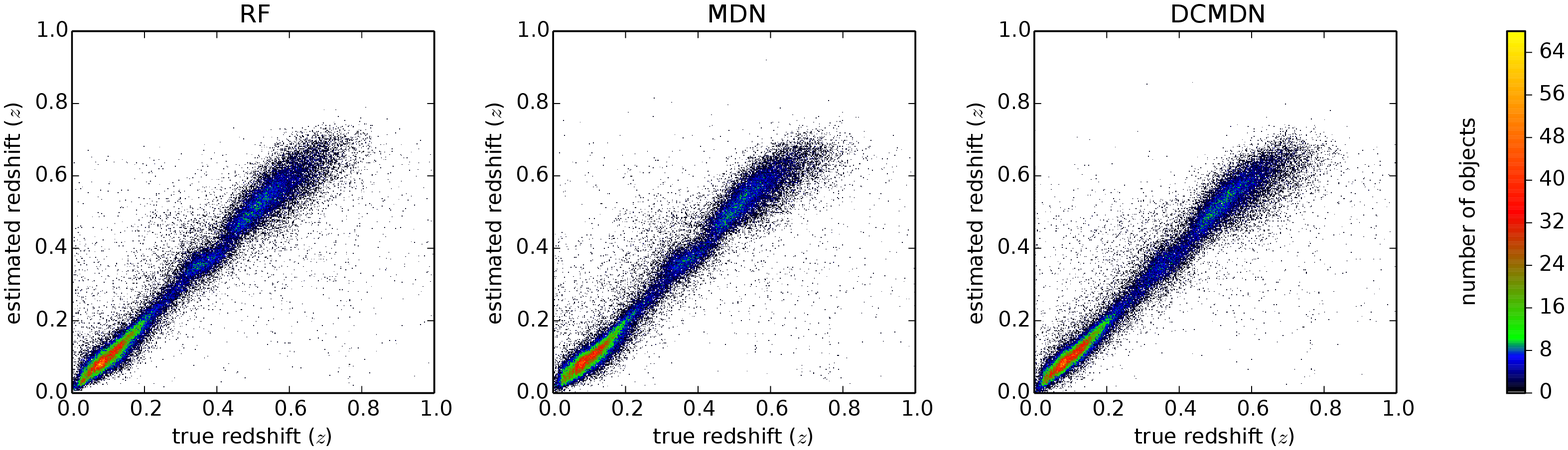}
  
  \vspace{3mm}
  {\footnotesize \hspace{3.0em} 1st mode, all $100,000$ objects \hspace{5.8em} 1st mode, all $100,000$ objects \hspace{5.5em} 1st mode, all $100,000$ objects \hfill\,}
  
  \includegraphics[width=\textwidth]{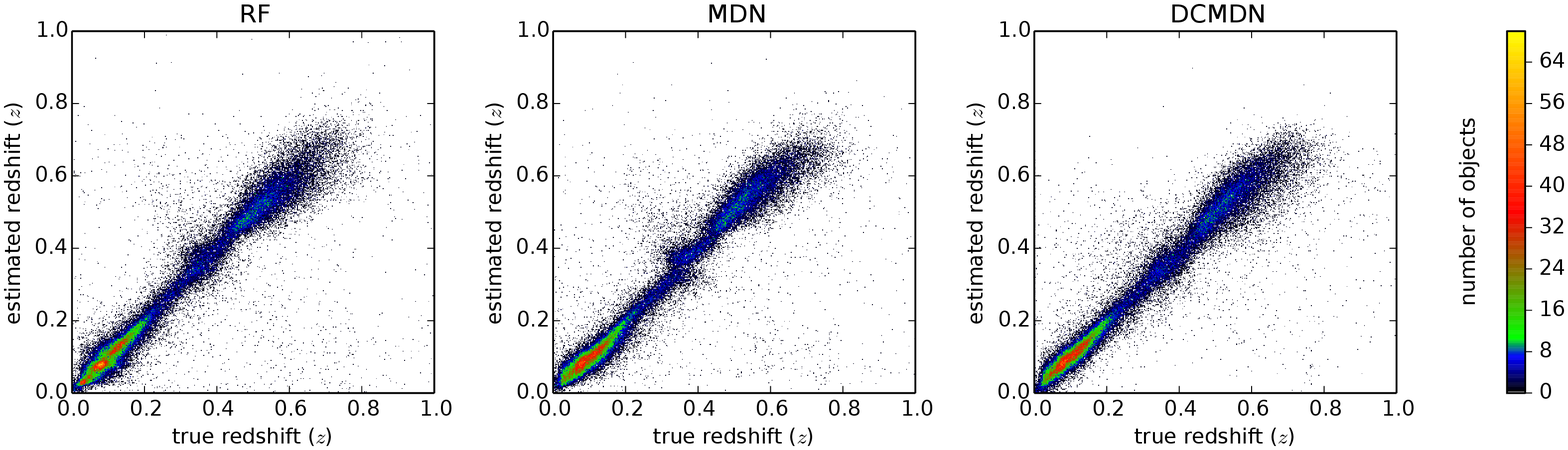}
  
  \vspace{3mm}
  {\footnotesize \hspace{3.0em} Mean, selected $65,327$ objects \hspace{5.5em} Mean, selected $87,284$ objects \hspace{5.2em} Mean, selected $83,148$ objects \hfill\,}
  
  \includegraphics[width=\textwidth]{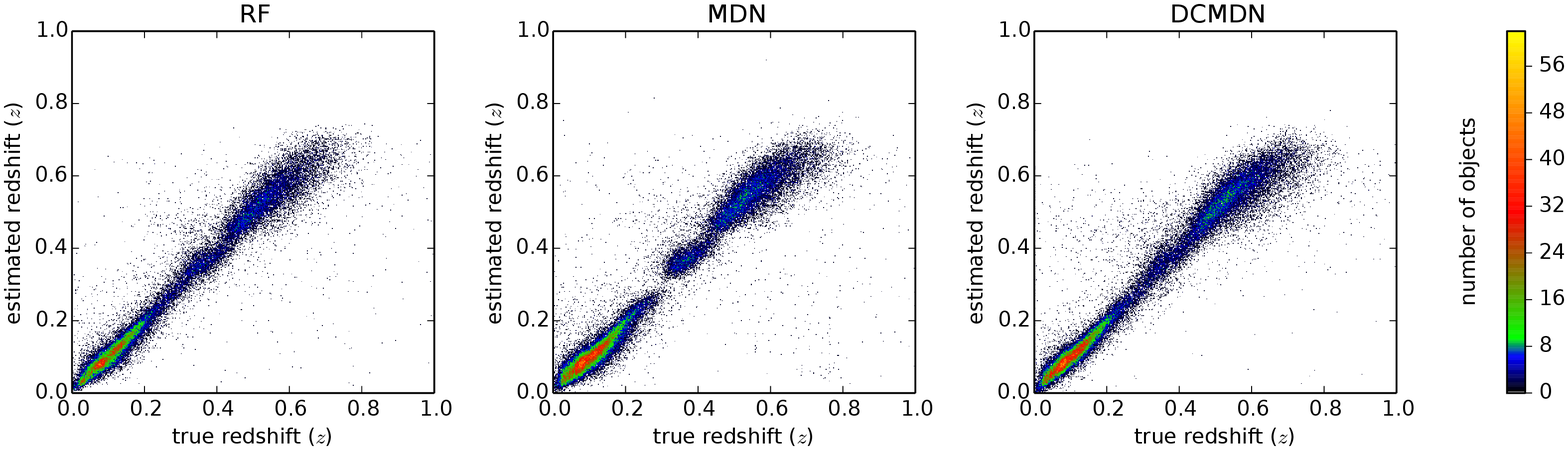}
    \caption{Results of Experiment 1 (galaxies).
    The estimated redshifts are plotted against the spectroscopic redshifts, which are considered ground trues.
    As in Table \ref{tab:galaxies_results} the plots are done for all three considered models (left to right) as well as with the mean, dominant mode and unambiguous objects (top to bottom).
    In addition, the number of unambiguous objects is reported.
    }
    \label{Fig:experiment_one_result}
\end{figure*}

\begin{table*}
  \centering
  \begin{tabular}{| c c | c | c | c | c | c | c | c | c | c  }
  \hline
  Criterion & Model & bias($\Delta z$) & $\sigma^{2}$($\Delta z$) & mad($\Delta z$) & rmse($\Delta z$) & bias($\Delta z_{norm}$) & $\sigma^{2}$($\Delta z_{norm}$) & mad($\Delta z_{norm}$) & rmse($\Delta z_{norm}$)\\
  \hline
  
  & RF & 0.0001 & 0.0033 & 0.0164 & 0.0575 & -0.0017 & 0.0018 & 0.0133 & 0.0431\\
  mean & MDN & 0.0016 & 0.0034 & 0.0174 & 0.0589 & -0.0003 & 0.0019 & 0.0141 & 0.0442\\
  & DCMDN & 0.0018 & 0.0030 & 0.0157 & 0.0548 & -0.0003 & 0.0017 & 0.0128 & 0.0409\\\hline
  
  & RF & 0.0031 & 0.0042 & 0.0171 & 0.0652 & 0.0010 & 0.0022 & 0.0139 & 0.0471\\
  first mode & MDN & 0.0029 & 0.0039 & 0.0172 & 0.0628 & 0.0010 & 0.0021 & 0.0140 & 0.0459\\
  & DCMDN & 0.0060 & 0.0031 & 0.0167 & 0.0561 & 0.0029 & 0.0016 & 0.0135 & 0.0407\\\hline
  
  & RF & 0.0001 & 0.0023 & 0.0147 & 0.0484 & -0.0011 & 0.0013 & 0.0121 & 0.0365\\
  selected & MDN & 0.0021 & 0.0027 & 0.0165 & 0.0523 & 0.0006 & 0.0015 & 0.0136 & 0.0391\\
  & DCMDN & 0.0017 & 0.0023 & 0.0146 & 0.0485 & -0.0001 & 0.0013 & 0.0120 & 0.0366\\\hline
 
  \end{tabular}
  \caption{Results of Experiment 1 (galaxies).
  Based on the estimated PDFs the traditional scores have been calculated.
  This was done by using the mean, the most dominant mode and the mean of the selected unambiguous objects, respectively.}
\label{tab:galaxies_results}
\end{table*}

\begin{figure*}
\centering
  \includegraphics[width=1.0\textwidth]{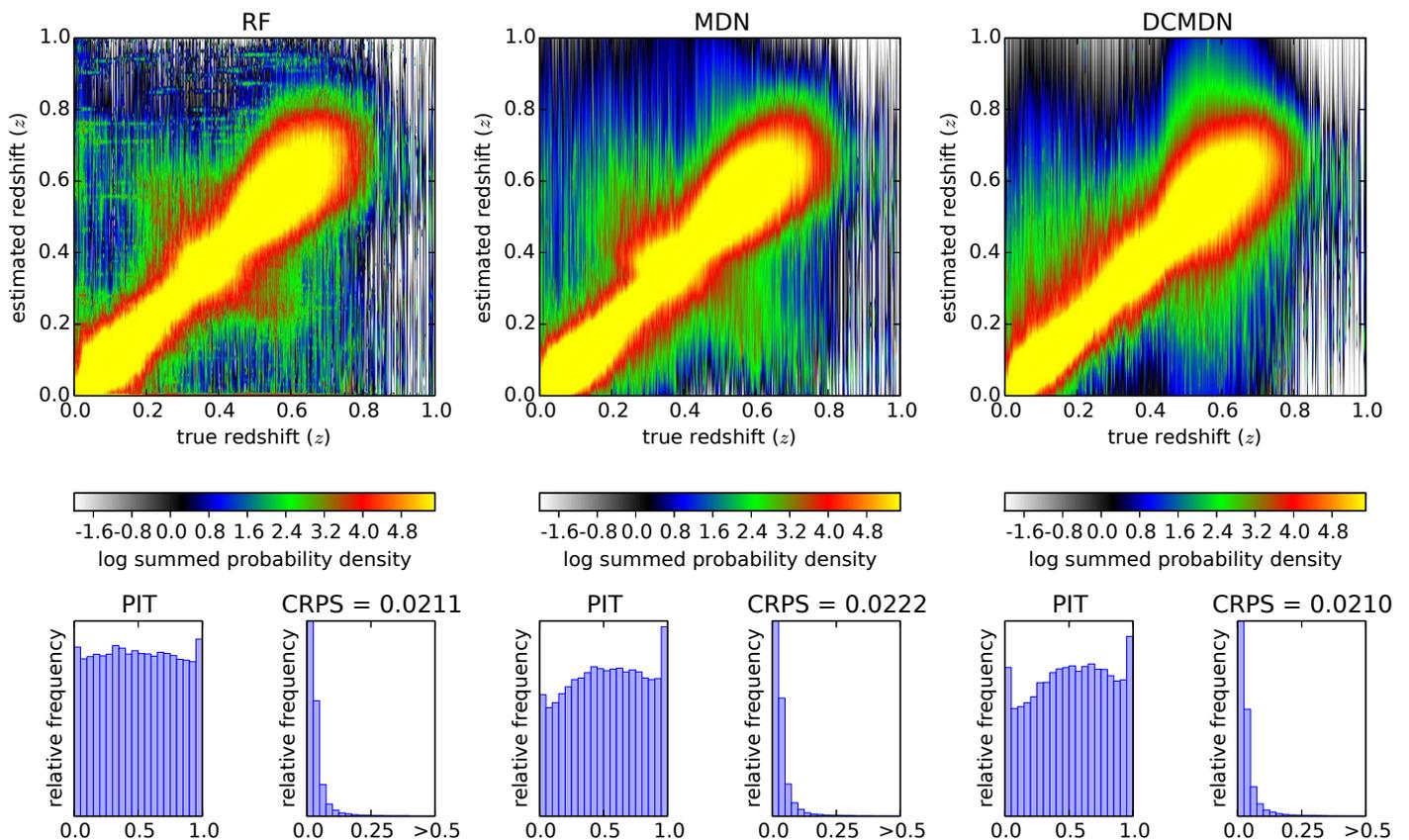}
    \caption{Results of Experiment 1 (galaxies) with a fully probabilistic evaluation and representation of the estimated PDFs.
    For each model, three plots are present.
    In the upper one, the predicted density distribution for each individual object is plotted at the corresponding spectroscopic redshift.
    The colors hereby indicate the logarithm of the summed probability densities using 500 redshift bins per axis.
    In the two lower plots, the histogram of the PIT values and the histogram of the CRPS values are shown, respectively.
    The mean CRPS values are reported, too.}
    \label{Fig:gal_crps_result}
\end{figure*}

\begin{figure*}[p!]
  {\footnotesize \hspace{3.2em} Mean, all $85,000$ objects \hspace{8.0em} Mean, all $85,000$ objects \hspace{7.8em} Mean, all $85,000$ objects \hfill\,}
  \includegraphics[width=\textwidth]{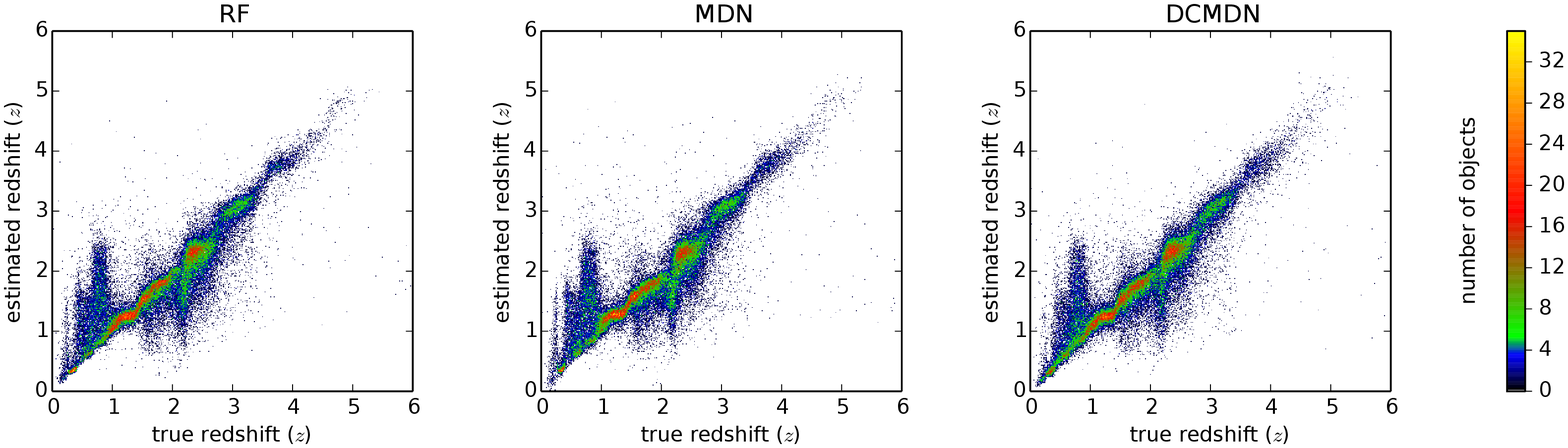}
    
  \vspace{3mm}
  {\footnotesize \hspace{2.7em} 1st mode, all $85,000$ objects \hspace{6.5em} 1st mode, all $85,000$ objects \hspace{6.2em} 1st mode, all $85,000$ objects \hfill\,}
  
  \includegraphics[width=\textwidth]{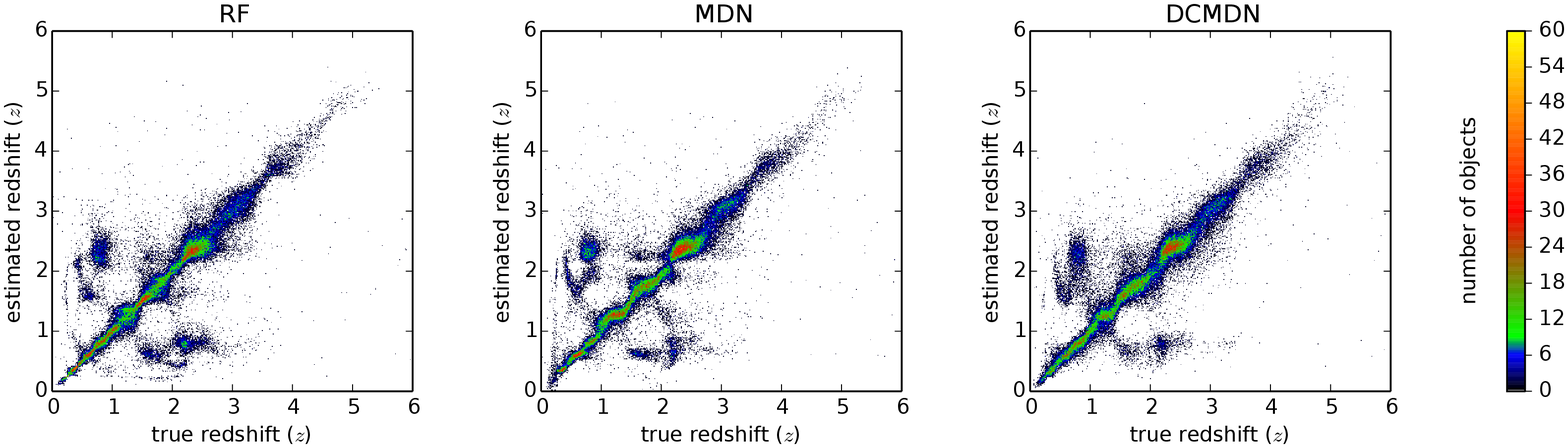}
    
  \vspace{3mm}
  {\footnotesize \hspace{2.3em} Mean, selected $39,087$ objects \hspace{5.5em} Mean, selected $43,916$ objects \hspace{5.7em} Mean, selected $40,558$ objects \hfill\,}
  \includegraphics[width=\textwidth]{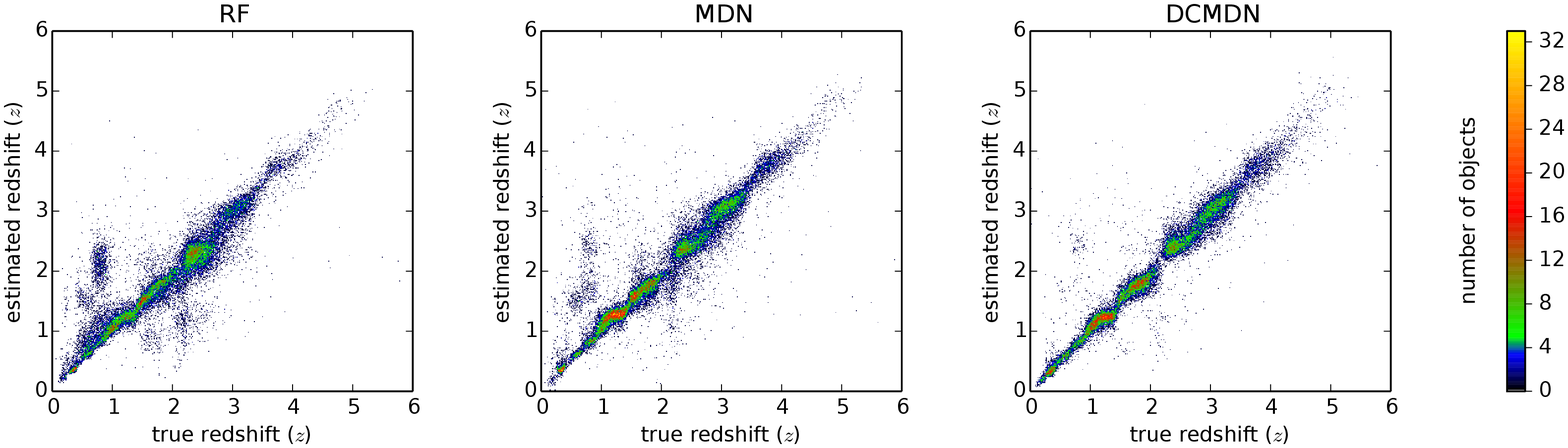}
    \caption{Results of Experiment 2 (quasars).
    The estimated redshifts are plotted against the spectroscopic redshifts.
    As in Fig.~\ref{Fig:experiment_one_result} the plots are sorted by the models (left to right) and the extracted point estimate (top to bottom).}
    \label{Fig:experiment_two_result}
\end{figure*}

\begin{table*}[p!]
  \centering
  \begin{tabular}{| c c | c | c | c | c | c | c | c | c | c  }
  \hline
  Criterion & Model & bias($\Delta z$) & $\sigma^{2}$($\Delta z$) & mad($\Delta z$) & rmse($\Delta z$) & bias($\Delta z_{norm}$) & $\sigma^{2}$($\Delta z_{norm}$) & mad($\Delta z_{norm}$) & rmse($\Delta z_{norm}$)\\
  \hline
  
  & RF & 0.007 & 0.217 & 0.145 & 0.466 & -0.033 & 0.048 & 0.050 & 0.222\\
  mean & MDN & -0.002 & 0.216 & 0.156 & 0.465 & -0.037 & 0.048 & 0.054 & 0.223\\
  & DCMDN & 0.011 & 0.168 & 0.128 & 0.411 & -0.023 & 0.035 & 0.045 & 0.189 \\\hline
  
  & RF & 0.002 & 0.319 & 0.087 & 0.564 & -0.026 & 0.066 & 0.031 & 0.258\\
  first mode & MDN & -0.058 & 0.282 & 0.095 & 0.535 & -0.052 & 0.068 & 0.034 & 0.267\\
  & DCMDN & -0.043 & 0.206 & 0.095 & 0.456 & -0.038 & 0.048 & 0.034 & 0.222\\\hline
  
  & RF & 0.005 & 0.162 & 0.111 & 0.402 & -0.024 & 0.037 & 0.039 & 0.194\\
  selected & MDN & -0.010 & 0.098 & 0.086 & 0.314 & -0.017 & 0.021 & 0.030 & 0.145\\
  & DCMDN & 0.004 & 0.047 & 0.075 & 0.217 & -0.004 & 0.009 & 0.026 & 0.095\\\hline
  
  \end{tabular}
  \caption{Results of Experiment 2 (quasars).
  Similarly to experiment 1, the traditional scores are presented.
  The score have been calculated with the mean, the most dominant mode and the mean of the selected unambiguous objects.}
\label{tab:quasar_results}
\end{table*} 

\begin{figure*}
\centering
  \includegraphics[width=1.0\textwidth]{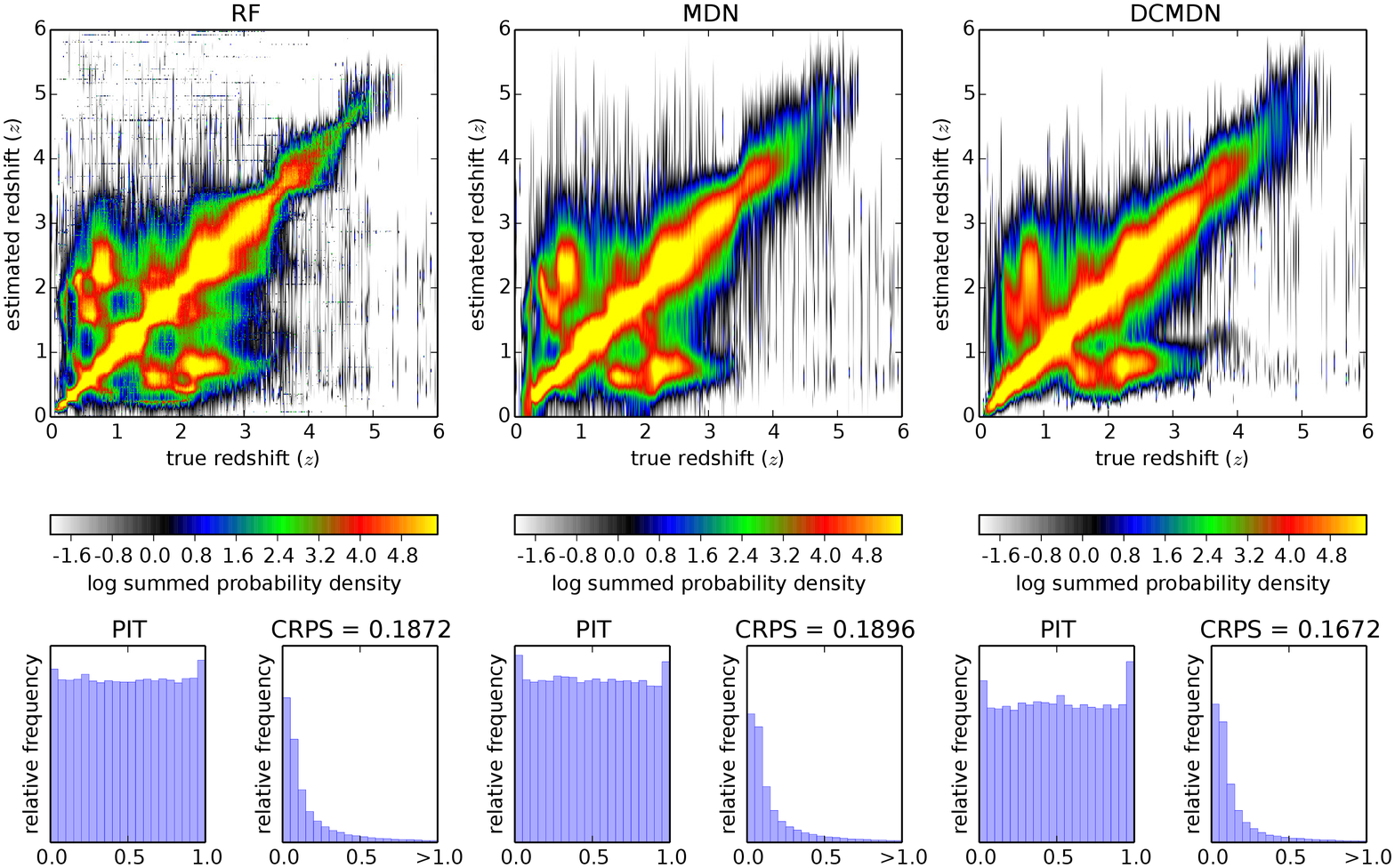}
    \caption{Results of Experiment 2 (quasars) with a fully probabilistic evaluation.
    As in Fig.~\ref{Fig:gal_crps_result} three plots per model are used to visualize the probabilistic performance of the estimated PDFs}
    \label{Fig:qso_crps_result}
\end{figure*}

The results of the first experiment are depicted in Figs.~\ref{Fig:experiment_one_result} and \ref{Fig:gal_crps_result}.
In both figures the estimated redshifts are plotted against the spectroscopic redshifts.
To be comparable to results from other publications in the field of photometric redshift estimation, in Fig.~\ref{Fig:experiment_one_result} the complex estimated PDFs are compressed into a single point estimate.
This compression is realized by either taking the plain mean, the first and most dominant mode of the mixture model or by taking the mean of objects that do not exhibit an ambiguous PDF (see Eq.~\ref{eq:constraints}).
Based on these three simplified representations of the estimated PDFs, the traditional scores have been calculated.
Those values are reported in Table \ref{tab:galaxies_results} and show a similar performance as other publications \citep[e.g.][]{2011mnras.418.2165l} that used a comparable data-set.
All three models used for testing show a very similar performance.
It is notable, that due to the nature of the used different models, the MDN and the DCMDN show a slightly better generalization performance especially in those regions where the transition of characteristic spectroscopic features through the broadband filters do not allow a distinct separation.
In those redshift regions where the degeneracy of the reverse determination of the spectral energy distribution and distinct spectral features through the low spectral resolution image data is dominating, selecting the first mode instead of using the mean value shows a poorer performance.
This is especially the case for $z\approx0.35$ and $z\approx0.45$.
When selecting only those objects for evaluation that do not have an ambiguous behavior, the mentioned redshift regions become underpopulated in the diagnostic plots.
For the DCMDN this effect is not as prominent as for the other two reference models due to the ability to make use of a larger base of information.
As presented in Fig.~\ref{Fig:experiment_one_result}, compressing the PDFs into single values does not cover the full complexity of the redshift estimations.
In particular, the selection of outliers having no unique single dominant redshift component in the PDF demonstrates the multi-modal nature of the photometric redshift estimation task.
Therefore, a visible improvement of the performance can be noticed when selecting only the subset of patterns which show no multi-modal behavior.
For these reasons, a proper probabilistic evaluation of the PDFs has to be performed.
Thus, in Fig.~\ref{Fig:gal_crps_result} a diagnostic plot is introduced which preserves the overall information of the density distributions.
Alongside this probabilistic comparison between the estimated redshift distributions and the spectroscopic values, the PIT and the CRPS are used as proper tools.
In the upper plot, the spectroscopic redshift is compared with the generated predictive density of every data item.
Hereby the logarithm of the summed probability density for each redshift bin is plotted.
In the two lower plots, the histograms for the PIT and the individual CRPS are given, together with the value of the mean CRPS. 
Analyzing the plots, it is notable that the RF performs slightly better with respect to the PIT.
In all cases the PIT shows a more or less well calibrated uniform distribution of the evaluated CDFs at the corresponding spectroscopic redshifts.
The MDN and the DCMDN in particular exhibit a more uniform and cleaner alignment toward the diagonal line which indicates the ideal performance.

\subsection{Experiment 2 - Quasars}

\begin{figure*}[p!]
  {\footnotesize \hspace{3.0em} Mean, all $285,000$ objects \hspace{7.1em} Mean, all $285,000$ objects \hspace{7.1em} Mean, all $285,000$ objects \hfill\,}
  \includegraphics[width=\textwidth]{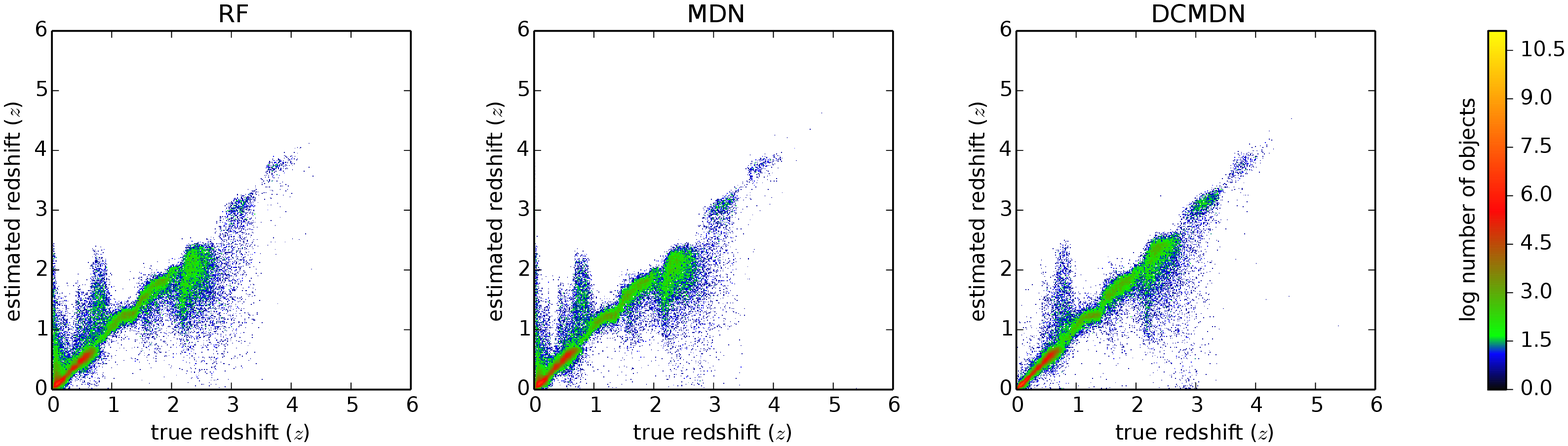}
    
  \vspace{6mm}
  {\footnotesize \hspace{2.2em} 1st mode, all $285,000$ objects \hspace{5.8em} 1st mode, all $285,000$ objects \hspace{5.5em} 1st mode, all $285,000$ objects \hfill\,}
  
  \includegraphics[width=\textwidth]{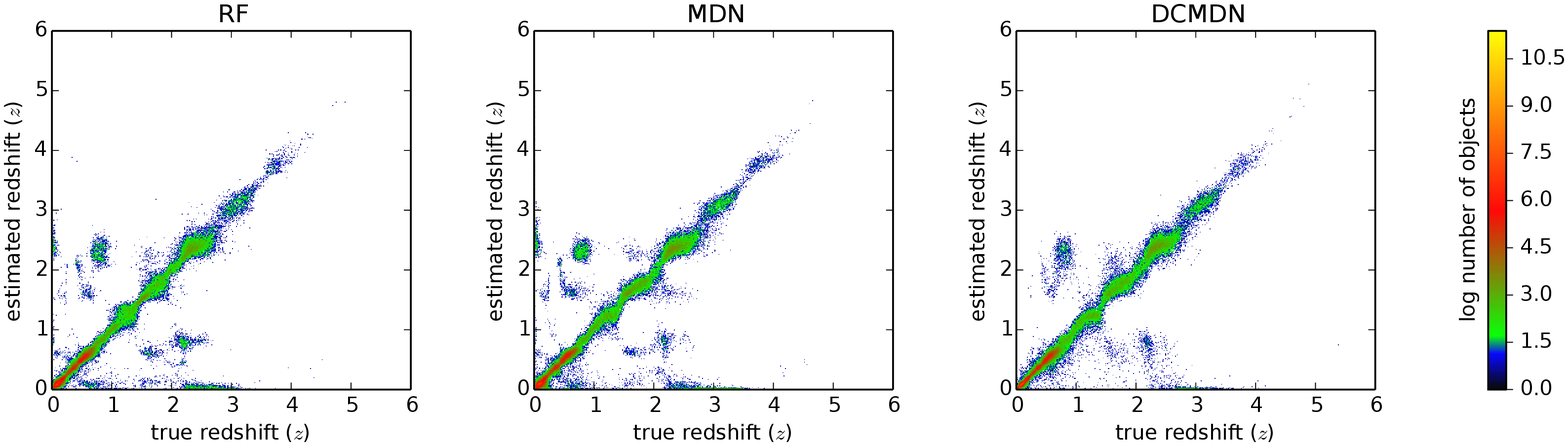}
    
  \vspace{6mm}
  {\footnotesize \hspace{2.0em} Mean, selected $170,923$ objects \hspace{4.8em} Mean, selected $175,362$ objects \hspace{4.8em} Mean, selected $226,041$ objects \hfill\,}
  
  \includegraphics[width=\textwidth]{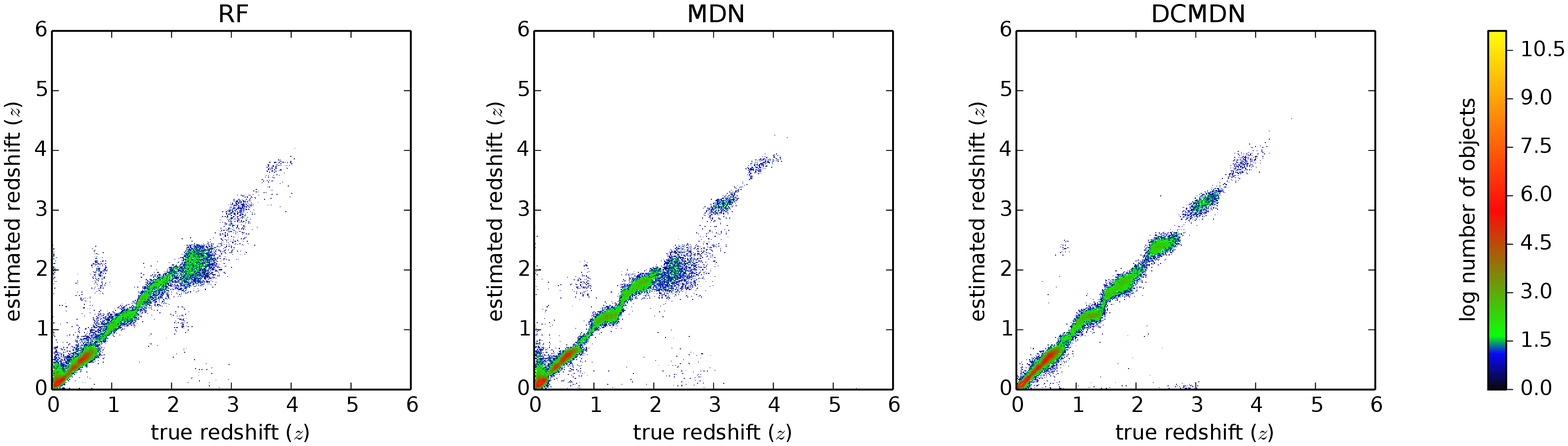}
    \caption{Results of Experiment 3 (mixed).
    The estimated redshifts are plotted against the spectroscopic redshifts.
    As in the figures of the two other experiments, the plots are sorted by the models (left to right) and the extracted point estimate (top to bottom).}
    \label{Fig:mixed_mean_result}
\end{figure*}

\begin{table*}[p!]
  \centering
  \begin{tabular}{| c c | c | c | c | c | c | c | c | c | c  }
  \hline
  Criterion & Model & bias($\Delta z$) & $\sigma^{2}$($\Delta z$) & mad($\Delta z$) & rmse($\Delta z$) & bias($\Delta z_{norm}$) & $\sigma^{2}$($\Delta z_{norm}$) & mad($\Delta z_{norm}$) & rmse($\Delta z_{norm}$)\\
  \hline
  
  & RF & -0.001 & 0.288 & 0.043 & 0.536 & -0.072 & 0.126 & 0.028 & 0.363\\
  mean & MDN & -0.006 & 0.279 & 0.043 & 0.528 & -0.073 & 0.125 & 0.031 & 0.362\\
  & DCMDN & 0.007 & 0.210 & 0.022 & 0.458 & -0.041 & 0.089 & 0.016 & 0.301\\\hline
  
  & RF & 0.040 & 0.435 & 0.020 & 0.660 & -0.029 & 0.150 & 0.013 & 0.388\\
  first mode & MDN & -0.027 & 0.393 & 0.024 & 0.627 & -0.067 & 0.191 & 0.016 & 0.442\\
  & DCMDN & -0.001 & 0.287 & 0.018 & 0.536 & -0.036 & 0.124 & 0.013 & 0.355\\\hline
  
  & RF & -0.001 & 0.118 & 0.016 & 0.343 & -0.029 & 0.059 & 0.012 & 0.245\\
  selected & MDN & -0.006 & 0.114 & 0.023 & 0.337 & -0.034 & 0.052 & 0.017 & 0.230\\
  & DCMDN & 0.002 & 0.043 & 0.012 & 0.206 & -0.007 & 0.014 & 0.010 & 0.120\\\hline
  
  \end{tabular}
  \caption{Results of Experiment 3 (mixed).
  Similarly to experiment 1 and 2, the traditional scores have been calculated.}
\label{tab:mixed_results}
\end{table*}

\begin{figure*}[t!]
  \includegraphics[width=\textwidth]{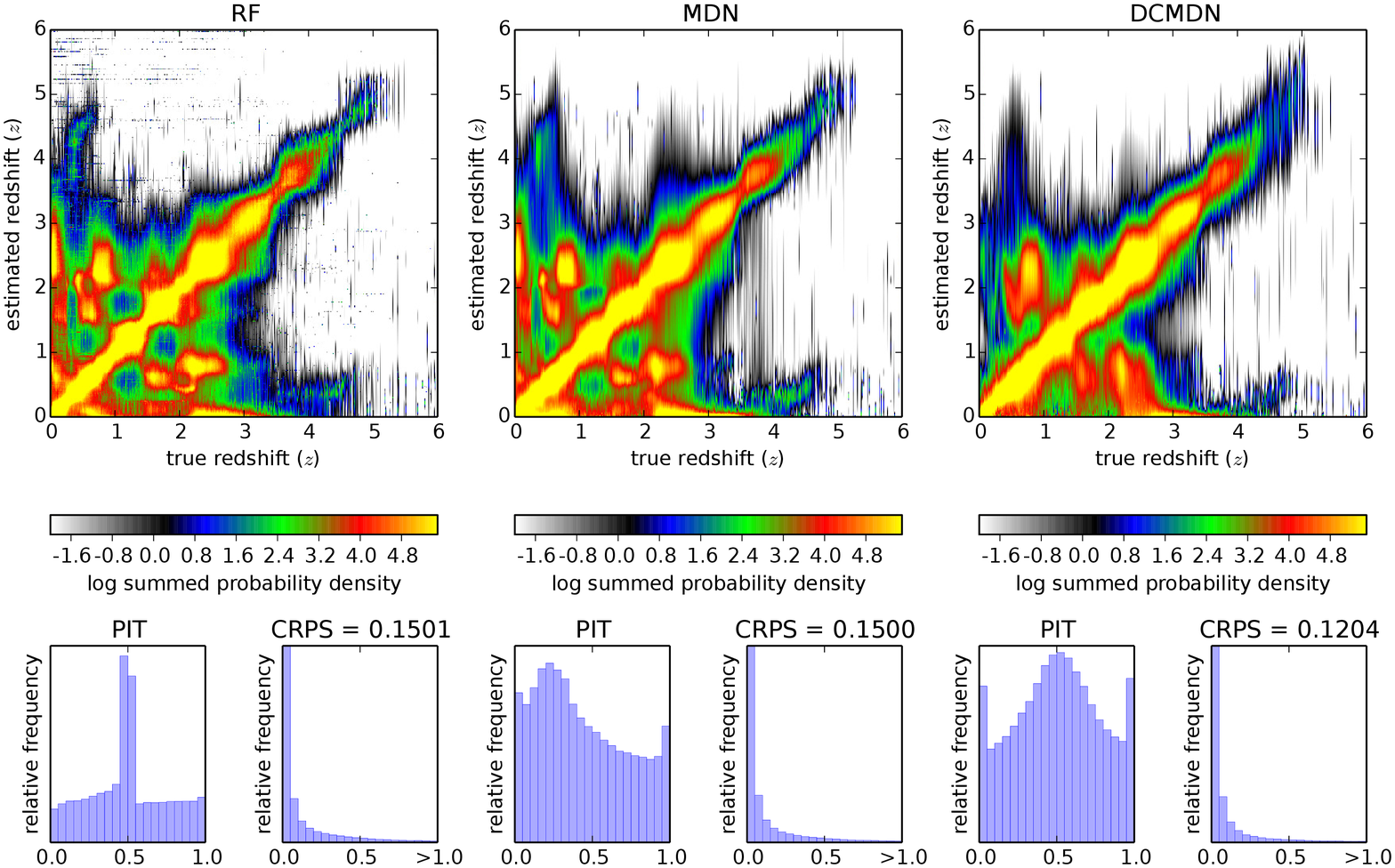}
    \caption{Results of Experiment 3 (mixed) with a fully probabilistic evaluation.
    As in Fig.~\ref{Fig:gal_crps_result} three plots per model are used to visualize the probabilistic performance of the estimated PDFs.}
    \label{Fig:mixed_crps_result}
\end{figure*}

In Fig.~\ref{Fig:experiment_two_result} the results of the second experiment are shown in the standard plot.
All estimated PDFs have been transferred into point estimates as it was done for the results of the first experiment.
In Table \ref{tab:quasar_results} the corresponding scores are presented.
As expected, the performances achieved with the mean values are comparable to point estimates obtained by other methods \citep{2011mnras.418.2165l}.
When using the first modes of the predicted density distributions, the degeneracies caused by the poor spectral resolution of the broadband photometry become visible.
Especially in the redshift areas around $z\approx[0.5..0.9]$ and $z\approx[1.5..2.5]$ a strong multi-modal behavior can be observed.
In those areas, picking the first mode does not necessary provide the best estimate, as both redshifts are equally likely.
The task of uniquely assigning a redshift based on poorly resolved broadband observations is in some regions a degenerated problem.
The better description of the probability distributions results therefore in a symmetry along the line of ideal performance.
Due to selection effects by observing a limited and cone-shaped volume, the distribution of objects with respect to cosmological scales results in not perfect symmetric behavior.
The necessity of a multi-modal description becomes even more obvious when objects with an ambiguous density distribution are not taken into consideration.
This way of excluding sources with estimated ambiguous redshift distributions can be considered a correct method to filter possible outliers, hence the selection of objects that show an uni-modal behavior is purely done based on the probabilistic description.
In all three different representation of the complex PDFs through a single point estimate, the DCMDN exhibits a better performance.
This is also reflected by the scores that are presented in Table \ref{tab:quasar_results}.
For the RF, the partitioning of the high-dimensional feature space orthogonal to the dimension axis does not provide a generalized representation of the regression problem.
When covering a wider redshift range the predictions exhibit more differences performance wise.
Therefore the performance of the RF drops in this experiment with respect to the other models as the MDN and DCMDN produce much smoother predictions along the ideal diagonal of the used diagnostic plot.
With respect to catastrophic outliers, the DCMDN has a superior performance when compared with the other two models.
This is consistent with the results from the first experiment.
When analyzing the probabilistic representation of the prediction results in Fig~\ref{Fig:qso_crps_result}, the superior performance of the DCMDN is striking.
The CRPS especially indicates the quality of the predictions made by the DCMDN.
In fact, by using images the DCMDN can utilize all the contained information and automatically extract the best usable features.

\subsection{Experiment 3 - Mixed}

The experiment with the mixed catalog is the most challenging one.
It tests how well the advantages of a deep convolutional network architecture can be used to render the step of pre-processing and pre-classifying objects obsolete.
This experiment is much closer to the real application compared to the previous cases.
In Fig.~\ref{Fig:mixed_mean_result} and Table \ref{tab:mixed_results} the results of Experiment~3 are reported and evaluated as single point estimates.
As in the previous experiments, the PDFs are therefore compressed.
When using the mean or the first mode as representation, the DCMDN significantly outperforms the RF and the MDN.
In the case of selecting unambiguous PDFs only, the result of the DCMDN is close to the ideal performance.
The obtained results confirm the indications of the previous experiments.
The fully probabilistic representation of the results of the third experiment are presented in Fig.~\ref{Fig:mixed_crps_result}.
When using proper tools and scores for evaluation, the DCMDN is the best model with respect to the CRPS.
Both, the RF and the MDN exhibit similar CRPS results, as they did in all the three experiments.
In the representation of the PITs the differences in calibration can be further analyzed.
The RF shows a nearly uniform distribution with an extreme peak in the center.
Due to the partitioning of the high-dimensional feature space performed by the RF, most of the stars are perfectly recovered.
As stars are fixedly assigned a redshift of $z=0$, the chosen way of fitting a GMM to the individual decision tree results produces the central peak.
This can be seen as an extreme overdispersion of the predictions in relation to the true values.
With respect to the estimated PDFs, the CDFs at the truncated true value $z=0$ is always very close to $0.5$.
For a stellar object, only a very few decision trees might return a redshift $z>0$, given the large number of $\approx 100,000$ stars used in the training sample.
During training the MDN got biased by the large number of stellar components with redshift set to $z=0$.
Therefore the corresponding PIT indicates a tendency to underestimate the redshift.
A complex, asymmetric behavior in the shape of the estimated PDFs is not captured when using the classical scores and diagnostic plots only, as the bias on the mean of the PDFs is close to zero for this experiment.
For the DCMDN we can observe two short comings.
Overall, the generated PDFs are overdispersed.
This effect is caused by the large fraction of stellar objects with a fixed value and the other objects covering the full redshift range.
The model tries to account for this highly unbalanced distribution in the redshift space when being optimized.
Generating broader distributions for one part of the objects while creating very narrow estimates for the stellar population is hard to be achieved by a generalizing model.
The second effect observed is the presence of outliers where the CDF indicates an extreme value.
This is caused by objects in the high redshift range being highly underrepresented in the training sample.
The DCMDN architecture has a superior capability of generalization and makes better use of the information contained in the original images, giving a boost in the final performance, together with a good separation of the stars from quasars.

\section{Conclusions}\label{sec:conclusions}

The aim of this work is to present and test a new method for photometric redshift estimation.
The novelty of the presented approach is to estimate probability density functions for redshifts based on imaging data.
The final goal is to make the additional steps of feature-extraction and feature-selection obsolete.
To achieve this, a deep convolutional network architecture was combined with a mixture density network.
Essential for a proper training is the use of the CRPS as loss function, taking into account not only the location but also the shape and layout of the estimated density distributions.
The new architecture is described in a general and conceptional way that allows using the concept for many other regression tasks in astronomy.

In order to perform a fair evaluation of the performance of the proposed method, three experiments have been performed.
Three different catalogs have been utilized for evaluation, containing galaxies, quasars and a mix of the previous catalogs plus a sample of stars.
The experiments were chosen to test the performance on different redshift ranges and different sources.
The last experiment was designed to test the model in a more realistic scenario, where a contamination with stellar sources and therefore a confusion between no and high redshift objects is synthetically introduced.
This experiment is intended to test whether the pre-classification of objects can be omitted, too.

In all three experiments a modified version of the random forest and a mixture density network are used as feature-based reference models.
To be comparable to the literature, the traditional scores and diagnostic plots have been used.
As we demonstrate, the usual way of expressing the results of the prediction quality through the traditional scores is not able to capture the complexity of often multi-modal and asymmetric distributions that are required to correctly describe the redshift estimates.
Therefore, proper scores and tools have been applied in addition to analyzing the results in a probabilistic way, that is, the CRPS and PIT.
These indicators can be considered proper tools to estimate the quality of photometric redshift density distributions.
As summarized in Table \ref{tab:summary_crps_table}, the DCMDN architecture outperforms the two reference models.
The same relative performances can be observed with the traditional diagnostics, too.
When using the PDFs to find and exclude ambiguous objects, the DCMDN always produces the best results.
Its ability to generalize from the training data, together with the larger amount of used available information permits a better probabilistic estimate of the redshift distribution and therefore a better selection of spurious predictions.

\begin{table}[b]
  \centering
  \begin{tabular}{| c c | c | c |}
  \hline
  Exp. & Model & CRPS & PIT\\
  \hline
  
  & RF & 0.021 & well calibrated\\
  Galaxies & MDN & 0.022 & biased\\
  & DCMDN & 0.021 & slightly overdispersed\\\hline
  
  & RF & 0.187 & well calibrated\\
  Quasars & MDN & 0.190 & well calibrated, few outliers\\
  & DCMDN & 0.167 & well calibrated, few outliers\\\hline
  
  & RF & 0.150 & extremely overdispersed\\
  Mixed & MDN & 0.150 & overdispersed and biased\\
  & DCMDN & 0.120 & overdispersed, some outliers\\\hline
  
  \end{tabular}
  \caption{Summary table of all the experiments}
\label{tab:summary_crps_table}
\end{table}

As shown in the last experiment, the DCMDN performs two very important tasks automatically.
During the training of the network, a set of thousands of features is automatically extracted and selected from the imaging data.
This minimizes the biases that are introduced when manually extracting and selecting features and increases the amount of utilized information.
The second and probably most important capability of the DCMDN is to solve the regression problem without the necessity of pre-classifying the objects.
The estimated PDFs reflect in the distribution of the probabilities the uncertainties of the classification as well as the uncertainties of the redshift estimation.
This is extremely important when dealing with data from larger surveys.
The errors introduced through a hard classification into distinct classes limits the ability of finding rare but interesting objects, like high redshifted quasars that are easily mistaken with cool faint stars.
A wrong initial classification would mark those quasars as stellar components even though the probability of being a high redshifted object is not negligible.
A fully probabilistic approach including feature extraction, feature selection and source classification is less affected by selection biases.

The performance of the two feature-based reference methods is in accordance with results from the literature.
The boost in performance observed for the DCMDN is related to the better use of information and the automatic selection of the best performing features.
As the DCMDN model was trained with data from the SDSS it could in principle be applied to every source in the SDSS database, without concern for the nature of the source, by directly using the images.
Only the selection biases that have been introduced when selecting the targets for spectroscopic follow-up observations in SDSS have to be considered, as those are preserved by the trained model.
Our approach provides the machinery to deal with the avalanche of data we are facing with the new generation of survey instruments already today.
Such fully-automatized and probabilistic approaches, based on deep learning architectures, are therefore necessary.

\begin{acknowledgements}
The authors gratefully acknowledge the support of the Klaus Tschira Foundation.

We gratefully thank the anonymous referee for the constructive and precise comments, leading to a more detailed discussion of the loss functions and the number of Gaussian components used.

We would like to thank Nikos Gianniotis and Erica Hopkins for proofreading and commenting on this work.

This work is based on the usage of Theano \citep{2016arXiv160502688short} to run our experiments.

Topcat \citep{2005aspc..347...29t} was used to access and visualize data from the VO and to create the used catalogs.

This work is based on data provided by SDSS.
Funding for SDSS-III has been provided by the Alfred P. Sloan Foundation, the Participating Institutions, the National Science Foundation, and the U.S. Department of Energy Office of Science. The SDSS-III web site is http://www.sdss3.org/.
SDSS-III is managed by the Astrophysical Research Consortium for the Participating Institutions of the SDSS-III Collaboration including the University of Arizona, the Brazilian Participation Group, Brookhaven National Laboratory, Carnegie Mellon University, University of Florida, the French Participation Group, the German Participation Group, Harvard University, the Instituto de Astrofisica de Canarias, the Michigan State/Notre Dame/JINA Participation Group, Johns Hopkins University, Lawrence Berkeley National Laboratory, Max Planck Institute for Astrophysics, Max Planck Institute for Extraterrestrial Physics, New Mexico State University, New York University, Ohio State University, Pennsylvania State University, University of Portsmouth, Princeton University, the Spanish Participation Group, University of Tokyo, University of Utah, Vanderbilt University, University of Virginia, University of Washington, and Yale University.

\end{acknowledgements}

\bibliographystyle{aa} 
\bibliography{references} 

\begin{thebibliography}{37}
\expandafter\ifx\csname natexlab\endcsname\relax\def\natexlab#1{#1}\fi

\bibitem[{Abdalla {et~al.}(2011)Abdalla, Banerji, Lahav, \&
  Rashkov}]{doi:10.1111/j.1365-2966.2011.19375.x}
Abdalla, F.~B., Banerji, M., Lahav, O., \& Rashkov, V. 2011, Monthly Notices of
  the Royal Astronomical Society, 417, 1891

\bibitem[{{Ahn} {et~al.}(2012){Ahn}, {Alexandroff}, {Allende Prieto},
  {Anderson}, {Anderton}, {Andrews}, {Aubourg}, {Bailey}, {Balbinot}, {Barnes},
  \& et~al.}]{2012ApJS..203...21A}
{Ahn}, C.~P., {Alexandroff}, R., {Allende Prieto}, C., {et~al.} 2012, \apjs,
  203, 21

\bibitem[{{Ball} \& {Brunner}(2010)}]{2010ijmpd..19.1049b}
{Ball}, N.~M. \& {Brunner}, R.~J. 2010, International Journal of Modern Physics
  D, 19, 1049

\bibitem[{{Beck} {et~al.}(2016){Beck}, {Dobos}, {Budav{\'a}ri}, {Szalay}, \&
  {Csabai}}]{2016MNRAS.460.1371B}
{Beck}, R., {Dobos}, L., {Budav{\'a}ri}, T., {Szalay}, A.~S., \& {Csabai}, I.
  2016, \mnras, 460, 1371

\bibitem[{Benítez(2000)}]{0004-637X-536-2-571}
Benítez, N. 2000, The Astrophysical Journal, 536, 571

\bibitem[{Bishop(1994)}]{Bishop94mixturedensity}
Bishop, C.~M. 1994, Mixture density networks, Tech. rep., Aston University

\bibitem[{Blake \& Bridle(2005)}]{doi:10.1111/j.1365-2966.2005.09526.x}
Blake, C. \& Bridle, S. 2005, Monthly Notices of the Royal Astronomical
  Society, 363, 1329

\bibitem[{{Bolzonella} {et~al.}(2000){Bolzonella}, {Miralles}, \&
  {Pell{\'o}}}]{2000A&A...363..476B}
{Bolzonella}, M., {Miralles}, J.-M., \& {Pell{\'o}}, R. 2000, \aap, 363, 476

\bibitem[{Breiman(2001)}]{Breiman:2001:RF:570181.570182}
Breiman, L. 2001, Mach. Learn., 45, 5

\bibitem[{Brescia {et~al.}(2013)Brescia, Cavuoti, D'Abrusco, Longo, \&
  Mercurio}]{Brescia2013}
Brescia, M., Cavuoti, S., D'Abrusco, R., Longo, G., \& Mercurio, A. 2013,
  Astrophysical Journal, 772

\bibitem[{Brescia {et~al.}(2014)Brescia, Cavuoti, Longo, \&
  De~Stefano}]{Brescia2014}
Brescia, M., Cavuoti, S., Longo, G., \& De~Stefano, V. 2014, Astronomy and
  Astrophysics, 568

\bibitem[{{Carliles} {et~al.}(2010){Carliles}, {Budav{\'a}ri}, {Heinis},
  {Priebe}, \& {Szalay}}]{2010ApJ...712..511C}
{Carliles}, S., {Budav{\'a}ri}, T., {Heinis}, S., {Priebe}, C., \& {Szalay},
  A.~S. 2010, ApJ, 712, 511

\bibitem[{{Carrasco Kind} \& {Brunner}(2013)}]{2013MNRAS.432.1483C}
{Carrasco Kind}, M. \& {Brunner}, R.~J. 2013, \mnras, 432, 1483

\bibitem[{Cavuoti {et~al.}(2015)Cavuoti, Brescia, Tortora, Longo, Napolitano,
  Radovich, La~Barbera, Capaccioli, de~Jong, Getman, Grado, \&
  Paolillo}]{Cavuoti20153100}
Cavuoti, S., Brescia, M., Tortora, C., {et~al.} 2015, Monthly Notices of the
  Royal Astronomical Society, 452, 3100

\bibitem[{{Collister} \& {Lahav}(2004)}]{2004PASP..116..345C}
{Collister}, A.~A. \& {Lahav}, O. 2004, \pasp, 116, 345

\bibitem[{{Dawson} {et~al.}(2013){Dawson}, {Schlegel}, {Ahn}, {Anderson},
  {Aubourg}, {Bailey}, {Barkhouser}, {Bautista}, {Beifiori}, {Berlind},
  {Bhardwaj}, {Bizyaev}, {Blake}, {Blanton}, {Blomqvist}, {Bolton}, {Borde},
  {Bovy}, {Brandt}, {Brewington}, {Brinkmann}, {Brown}, {Brownstein}, {Bundy},
  {Busca}, {Carithers}, {Carnero}, {Carr}, {Chen}, {Comparat}, {Connolly},
  {Cope}, {Croft}, {Cuesta}, {da Costa}, {Davenport}, {Delubac}, {de Putter},
  {Dhital}, {Ealet}, {Ebelke}, {Eisenstein}, {Escoffier}, {Fan}, {Filiz Ak},
  {Finley}, {Font-Ribera}, {G{\'e}nova-Santos}, {Gunn}, {Guo}, {Haggard},
  {Hall}, {Hamilton}, {Harris}, {Harris}, {Ho}, {Hogg}, {Holder}, {Honscheid},
  {Huehnerhoff}, {Jordan}, {Jordan}, {Kauffmann}, {Kazin}, {Kirkby}, {Klaene},
  {Kneib}, {Le Goff}, {Lee}, {Long}, {Loomis}, {Lundgren}, {Lupton}, {Maia},
  {Makler}, {Malanushenko}, {Malanushenko}, {Mandelbaum}, {Manera}, {Maraston},
  {Margala}, {Masters}, {McBride}, {McDonald}, {McGreer}, {McMahon}, {Mena},
  {Miralda-Escud{\'e}}, {Montero-Dorta}, {Montesano}, {Muna}, {Myers},
  {Naugle}, {Nichol}, {Noterdaeme}, {Nuza}, {Olmstead}, {Oravetz}, {Oravetz},
  {Owen}, {Padmanabhan}, {Palanque-Delabrouille}, {Pan}, {Parejko},
  {P{\^a}ris}, {Percival}, {P{\'e}rez-Fournon}, {P{\'e}rez-R{\`a}fols},
  {Petitjean}, {Pfaffenberger}, {Pforr}, {Pieri}, {Prada}, {Price-Whelan},
  {Raddick}, {Rebolo}, {Rich}, {Richards}, {Rockosi}, {Roe}, {Ross}, {Ross},
  {Rossi}, {Rubi{\~n}o-Martin}, {Samushia}, {S{\'a}nchez}, {Sayres}, {Schmidt},
  {Schneider}, {Sc{\'o}ccola}, {Seo}, {Shelden}, {Sheldon}, {Shen}, {Shu},
  {Slosar}, {Smee}, {Snedden}, {Stauffer}, {Steele}, {Strauss}, {Streblyanska},
  {Suzuki}, {Swanson}, {Tal}, {Tanaka}, {Thomas}, {Tinker}, {Tojeiro},
  {Tremonti}, {Vargas Maga{\~n}a}, {Verde}, {Viel}, {Wake}, {Watson}, {Weaver},
  {Weinberg}, {Weiner}, {West}, {White}, {Wood-Vasey}, {Yeche}, {Zehavi},
  {Zhao}, \& {Zheng}}]{2013AJ....145...10D}
{Dawson}, K.~S., {Schlegel}, D.~J., {Ahn}, C.~P., {et~al.} 2013, \aj, 145, 10

\bibitem[{D'Isanto \& Polsterer(2017)}]{disanto2017esann}
D'Isanto, A. \& Polsterer, K.~L. 2017, in {ESANN} 2017, 25th European Symposium
  on Artificial Neural Networks, Bruges, Belgium, April 26-28, 2017,
  Proceedings

\bibitem[{{Dubath} {et~al.}(2017){Dubath}, {Apostolakos}, {Bonchi}, {Belikov},
  {Brescia}, {Cavuoti}, {Capak}, {Coupon}, {Dabin}, {Degaudenzi}, {Desai},
  {Dubath}, {Fontana}, {Fotopoulou}, {Frailis}, {Galametz}, {Hoar}, {Holliman},
  {Hoyle}, {Hudelot}, {Ilbert}, {Kuemmel}, {Melchior}, {Mellier}, {Mohr},
  {Morisset}, {Paltani}, {Pello}, {Pilo}, {Polenta}, {Poncet}, {Saglia},
  {Salvato}, {Sauvage}, {Schefer}, {Serrano}, {Soldati}, {Tramacere},
  {Williams}, \& {Zacchei}}]{2017IAUS..325...73D}
{Dubath}, P., {Apostolakos}, N., {Bonchi}, A., {et~al.} 2017, in IAU Symposium,
  Vol. 325, IAU Symposium, 73--82

\bibitem[{{Fern{\'a}ndez-Soto} {et~al.}(2001){Fern{\'a}ndez-Soto}, {Lanzetta},
  {Chen}, {Pascarelle}, \& {Yahata}}]{2001ApJS..135...41F}
{Fern{\'a}ndez-Soto}, A., {Lanzetta}, K.~M., {Chen}, H.-W., {Pascarelle},
  S.~M., \& {Yahata}, N. 2001, \apjs, 135, 41

\bibitem[{{Fernique} {et~al.}(2015){Fernique}, {Allen}, {Boch}, {Oberto},
  {Pineau}, {Durand}, {Bot}, {Cambr{\'e}sy}, {Derriere}, {Genova}, \&
  {Bonnarel}}]{2015A&A...578A.114F}
{Fernique}, P., {Allen}, M.~G., {Boch}, T., {et~al.} 2015, A\&A, 578, A114

\bibitem[{{Gneiting} {et~al.}(2005){Gneiting}, {Raftery}, {Westveld}, \&
  {Goldman}}]{2005MWRv..133.1098G}
{Gneiting}, T., {Raftery}, A.~E., {Westveld}, A.~H., \& {Goldman}, T. 2005,
  Monthly Weather Review, 133, 1098

\bibitem[{{Grimit} {et~al.}(2006){Grimit}, {Gneiting}, {Berrocal}, \&
  {Johnson}}]{2006QJRMS.132.2925G}
{Grimit}, E.~P., {Gneiting}, T., {Berrocal}, V.~J., \& {Johnson}, N.~A. 2006,
  Quarterly Journal of the Royal Meteorological Society, 132, 2925

\bibitem[{{Hersbach}(2000)}]{2000WtFor..15..559H}
{Hersbach}, H. 2000, Weather and Forecasting, 15, 559

\bibitem[{{Hoyle}(2016)}]{2016A&C....16...34H}
{Hoyle}, B. 2016, Astronomy and Computing, 16, 34

\bibitem[{K{\"{u}}gler {et~al.}(2016)K{\"{u}}gler, Gianniotis, \&
  Polsterer}]{KuglerGP16}
K{\"{u}}gler, S.~D., Gianniotis, N., \& Polsterer, K.~L. 2016, in 2016 {IEEE}
  Symposium Series on Computational Intelligence, {SSCI} 2016, Athens, Greece,
  December 6-9, 2016, 1--8

\bibitem[{{Laureijs} {et~al.}(2012){Laureijs}, {Gondoin}, {Duvet}, {Saavedra
  Criado}, {Hoar}, {Amiaux}, {Augu{\`e}res}, {Cole}, {Cropper}, {Ealet},
  {Ferruit}, {Escudero Sanz}, {Jahnke}, {Kohley}, {Maciaszek}, {Mellier},
  {Oosterbroek}, {Pasian}, {Sauvage}, {Scaramella}, {Sirianni}, \&
  {Valenziano}}]{2012spie.8442e..0tl}
{Laureijs}, R., {Gondoin}, P., {Duvet}, L., {et~al.} 2012, in \procspie, Vol.
  8442, Space Telescopes and Instrumentation 2012: Optical, Infrared, and
  Millimeter Wave, 84420T

\bibitem[{{Laurino} {et~al.}(2011){Laurino}, {D'Abrusco}, {Longo}, \&
  {Riccio}}]{2011mnras.418.2165l}
{Laurino}, O., {D'Abrusco}, R., {Longo}, G., \& {Riccio}, G. 2011, \mnras, 418,
  2165

\bibitem[{LeCun {et~al.}(1998)LeCun, Bottou, Bengio, \& Haffner}]{lecun-98}
LeCun, Y., Bottou, L., Bengio, Y., \& Haffner, P. 1998, Proceedings of the
  IEEE, 86, 2278

\bibitem[{{P{\^a}ris} {et~al.}(2012){P{\^a}ris}, {Petitjean}, {Aubourg},
  {Bailey}, {Ross}, {Myers}, {Strauss}, {Anderson}, {Arnau}, {Bautista},
  {Bizyaev}, {Bolton}, {Bovy}, {Brandt}, {Brewington}, {Browstein}, {Busca},
  {Capellupo}, {Carithers}, {Croft}, {Dawson}, {Delubac}, {Ebelke},
  {Eisenstein}, {Engelke}, {Fan}, {Filiz Ak}, {Finley}, {Font-Ribera}, {Ge},
  {Gibson}, {Hall}, {Hamann}, {Hennawi}, {Ho}, {Hogg}, {Ivezi{\'c}}, {Jiang},
  {Kimball}, {Kirkby}, {Kirkpatrick}, {Lee}, {Le Goff}, {Lundgren}, {MacLeod},
  {Malanushenko}, {Malanushenko}, {Maraston}, {McGreer}, {McMahon},
  {Miralda-Escud{\'e}}, {Muna}, {Noterdaeme}, {Oravetz},
  {Palanque-Delabrouille}, {Pan}, {Perez-Fournon}, {Pieri}, {Richards},
  {Rollinde}, {Sheldon}, {Schlegel}, {Schneider}, {Slosar}, {Shelden}, {Shen},
  {Simmons}, {Snedden}, {Suzuki}, {Tinker}, {Viel}, {Weaver}, {Weinberg},
  {White}, {Wood-Vasey}, \& {Y{\`e}che}}]{2012A&A...548A..66P}
{P{\^a}ris}, I., {Petitjean}, P., {Aubourg}, {\'E}., {et~al.} 2012, \aap, 548,
  A66

\bibitem[{{Polsterer} {et~al.}(2013){Polsterer}, {Zinn}, \&
  {Gieseke}}]{2013mnras.428..226p}
{Polsterer}, K.~L., {Zinn}, P.-C., \& {Gieseke}, F. 2013, \mnras, 428, 226

\bibitem[{Rosenblatt(1962)}]{rosenblatt1962principles}
Rosenblatt, F. 1962, Principles of neurodynamics: perceptrons and the theory of
  brain mechanisms, Report (Cornell Aeronautical Laboratory) (Spartan Books)

\bibitem[{{Sadeh} {et~al.}(2016){Sadeh}, {Abdalla}, \&
  {Lahav}}]{2016PASP..128j4502S}
{Sadeh}, I., {Abdalla}, F.~B., \& {Lahav}, O. 2016, \pasp, 128, 104502

\bibitem[{Salvato {et~al.}(2009)Salvato, Hasinger, Ilbert, Zamorani, Brusa,
  Scoville, Rau, Capak, Arnouts, Aussel, Bolzonella, Buongiorno, Cappelluti,
  Caputi, Civano, Cook, Elvis, Gilli, Jahnke, Kartaltepe, Impey, Lamareille,
  Floch, Lilly, Mainieri, McCarthy, McCracken, Mignoli, Mobasher, Murayama,
  Sasaki, Sanders, Schiminovich, Shioya, Shopbell, Silverman, Smolčić,
  Surace, Taniguchi, Thompson, Trump, Urry, \& Zamojski}]{0004-637X-690-2-1250}
Salvato, M., Hasinger, G., Ilbert, O., {et~al.} 2009, The Astrophysical
  Journal, 690, 1250

\bibitem[{{Schneider} {et~al.}(2010){Schneider}, {Richards}, {Hall}, {Strauss},
  {Anderson}, {Boroson}, {Ross}, {Shen}, {Brandt}, {Fan}, {Inada}, {Jester},
  {Knapp}, {Krawczyk}, {Thakar}, {Vanden Berk}, {Voges}, {Yanny}, {York},
  {Bahcall}, {Bizyaev}, {Blanton}, {Brewington}, {Brinkmann}, {Eisenstein},
  {Frieman}, {Fukugita}, {Gray}, {Gunn}, {Hibon}, {Ivezi{\'c}}, {Kent}, {Kron},
  {Lee}, {Lupton}, {Malanushenko}, {Malanushenko}, {Oravetz}, {Pan}, {Pier},
  {Price}, {Saxe}, {Schlegel}, {Simmons}, {Snedden}, {SubbaRao}, {Szalay}, \&
  {Weinberg}}]{2010AJ....139.2360S}
{Schneider}, D.~P., {Richards}, G.~T., {Hall}, P.~B., {et~al.} 2010, \aj, 139,
  2360

\bibitem[{Schwarz(1978)}]{schwarz1978}
Schwarz, G. 1978, Ann. Statist., 6, 461

\bibitem[{{Taylor}(2005)}]{2005aspc..347...29t}
{Taylor}, M.~B. 2005, in Astronomical Society of the Pacific Conference Series,
  Vol. 347, Astronomical Data Analysis Software and Systems XIV, ed.
  P.~{Shopbell}, M.~{Britton}, \& R.~{Ebert}, 29

\bibitem[{{Theano Development Team}(2016)}]{2016arXiv160502688short}
{Theano Development Team}. 2016, arXiv e-prints, abs/1605.02688

\end{thebibliography}

\appendix

\section{CRPS and PIT}\label{sec:crps_and_pit}

\begin{figure}[b]
  \centering
  \includegraphics[width=1.00\columnwidth]{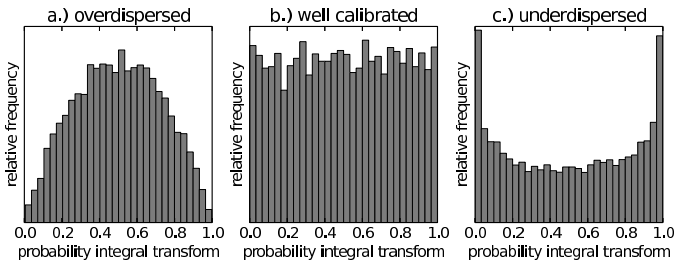}
    \caption{Visual guide to the usage of a PIT.
    In case the estimated PDFs are to broad with respect to the position of the true value, a convex histogram with a peak in the center can be observed (a).
    As soon as the predicted densities are too narrow, the evaluation of the CDF at the true redshift exhibits in most cases just very low and very high values.
    Therefore a concave, U-shaped histogram will be produced (c).
    Only in the case where the widths of the predicted densities is in accordance with the deviations from the true measurements, a uniformly distributed histogram is generated (b).
    This indicates sharp and well calibrated predictions.}
    \label{Fig:pit}
\end{figure}

The CRPS is defined by the relation:
\begin{equation}
CRPS = CRPS(F, x_{a}) =  \int_{- \infty}^{+ \infty}[F(x) - F_{a}(x)]^{2}dx
\end{equation}
\noindent where $F(x)$ and $F_{a}(x)$ are the CDFs relative to the predicted PDF $f(t)$ and the observation $x_{a}$, respectively. Namely: $F(x) = \int_{- \infty}^{x}f(t)dt$ and $F_{x} = H(x - x_{a})$, with $H(x)$ being the Heaviside step-function.
In case the PDF is given as a GMM, the CRPS can be calculated through the following formula in closed form:

\begin{equation}\begin{split}
CRPS\bigg( &\sum_{m=1}^{M}\omega_{m}\mathcal{N}(\mu_{m},\sigma_{m}^{2})\bigg) = \\
&\sum_{m=1}^{M}\omega_{m}A(x - \mu_{m}, \sigma_{m}^{2}) - \\
\frac{1}{2}&\sum_{m=1}^{M}\sum_{n=1}^{M}\omega_{m}\omega_{n}A(\mu_{m} - \mu_{n}, \sigma_{m}^{2} + \sigma_{n}^{2})
\end{split}
\label{GMMcrps}
\end{equation}

\noindent where

\begin{equation}
A(\mu, \sigma^{2}) = 2\sigma\phi\Big(\frac{\mu}{\sigma}\Big) + \mu\Big(2\Phi\Big(\frac{\mu}{\sigma}\Big) - 1\Big)
\end{equation}

\noindent and $\phi\Big(\frac{y - \mu}{\sigma}\Big)$, $\Phi\Big(\frac{y - \mu}{\sigma}\Big)$ respectively represent the PDF and the CDF of a normal distribution with a mean of zero and a variance of one evaluated through the normalized prediction error $\frac{y - \mu}{\sigma}$.

The probability integral transform (PIT) is generated with the histogram of the values:

\begin{equation}
p_{t} = F_{t}(x_{t})
\end{equation}

\noindent being $F_{t}$ the CDF of the predicted PDF evaluated at the observation $x_{t}$.
In Fig.~\ref{Fig:pit} some example PITs are given.

\vfill\null

\section{CRPS vs. log-likelihood as loss-function}\label{sec:CRPSvslL}

\begin{figure}[h]
  \centering
  \includegraphics[width=1.00\columnwidth]{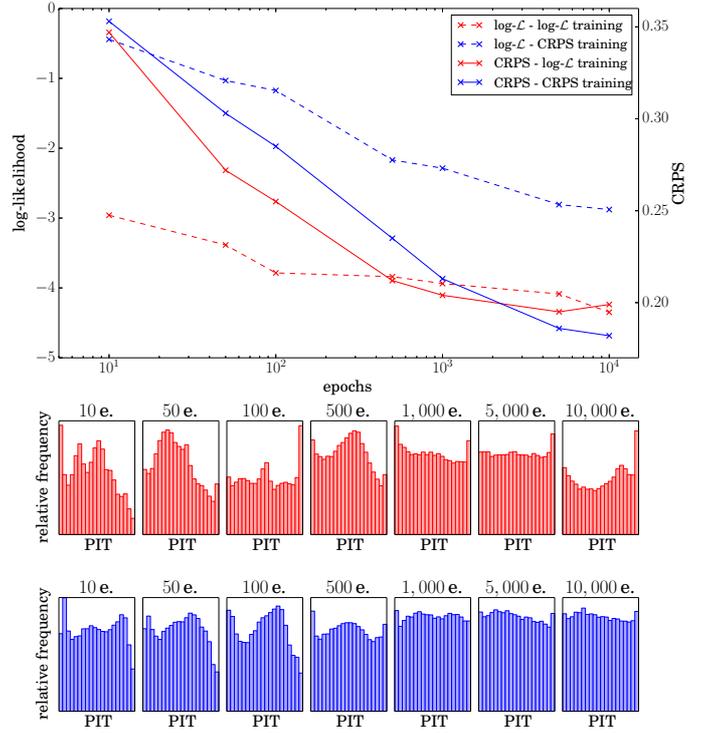}
    \caption{Comparison of the performance when using the log-likelihood (red) and the CRPS (blue) as loss-functions for training the MDN.
    For different epochs the CRPS and log-likelihood scores as well as the PIT histograms are provided.
    }
    \label{Fig:ll_train}
\end{figure}

We compared the performance of the MDN when trained with the log-likelihood and the CRPS as loss functions, in order to investigate the differences in the behavior of the model.
The experiment was performed on the data used for the quasar experiment, split in independent subsets for training ($100,000$\,objects) and testing ($85,000$\,objects).
The results are shown in Fig.~\ref{Fig:ll_train}.
In this plot we report the performance expressed by using both the log-likelihood and the CRPS as score functions, at different epochs of the training phase.
Moreover, the correspondent PIT histograms for these epochs are shown.
Training the network with the log-likelihood improves the performance in terms of the log-likelihood itself, respect to the same architecture trained using the CRPS.
As expected, when training with the CRPS, the observed results are opposite to the previous case.
The PIT histograms indicate a better performance when the neural network is trained using the CRPS, leading to a well calibrated PIT already after $500$ epochs.
Moreover, the PIT of the model trained using the log-likelihood starts to degrade again at $10,000$ epochs.
This does not happen when using the CRPS for training, as the CRPS accounts for calibration and sharpness of the predictions.
For this reason, the choice of the CRPS as loss function is reasonable, in order to obtain sharper and better calibrated PDFs.

\section{Number of Gaussian components}\label{sec:BIC}

\begin{figure}[h]
  \centering
  \includegraphics[width=1.00\columnwidth]{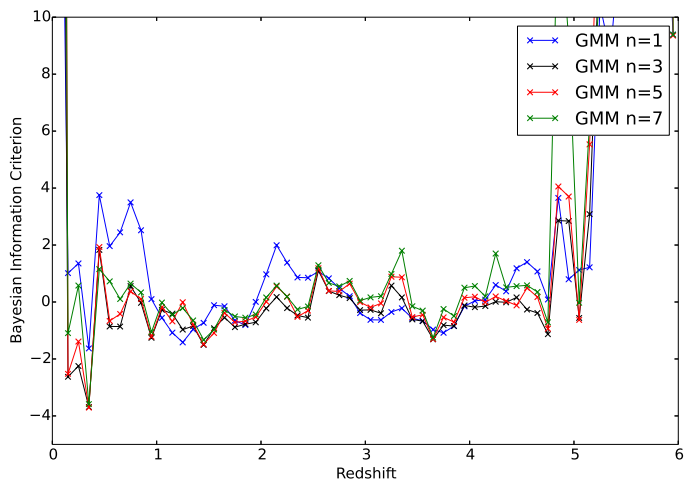}
    \caption{Distribution of the mean Bayesian information criterion score, with respect to redshift and number of Gaussian components.
    The plot depicts that different regions of the redshift range demand a different number of components.
    On average, the use of $n=5$ is a reasonable choice. The experiment is performed using the RF and the quasar data-set.}
    \label{Fig:bic}
\end{figure}

In order to choose an appropriate number of Gaussian components, an experiment has been performed using the RF model for the quasar data-set and calculating the BIC following \citet{schwarz1978}.
The mean score has been calculated over different redshift bins, for different numbers of Gaussian components.
The plot shows that extreme values, like $n=1$ or $n=7$ tend to exhibit bad performance in multiple regions.
Instead, the results using $n=3$ and $n=5$ are comparably good.
The BIC score is based on the log-likelihood calculation, therefore we consider a choice of $n=5$ to be reasonable for this work, keeping in mind that our model is trained using the CRPS. 

\section{Code}\label{sec:code}

The code used to do this research can be found on the ASCL: \url{http://www.ascl.net/ascl:1709.006}.

\section{Data}\label{sec:appdata}

The SDSS object IDs of the randomly extracted stars and galaxies are available as supplementary information.
In addition the SDSS IDs of the quasars are provided as a plain ASCII file.

The results of the predictions done with the DCMDN architecture for the three experiments are made available as ASCII files, too.
Those files contain the spectroscopic redshifts of the test objects followed by 15 outputs that can be used to calculate the GMM parameters (five means, five sigmas, five weights) as described in Eq.~\ref{eq:mu_sigma_omega}.
This allows reproduction of the performance of our model.

\vfill
\noindent
\textbf{\ttfamily galaxies\_id.csv} contains the SDSS object IDs of the galaxies used for the experiments.

\noindent
\textbf{\ttfamily quasars\_id.csv} contains the SDSS object IDs of the quasars used for the experiments.

\noindent
\textbf{\ttfamily stars\_id.csv} contains the SDSS object IDs of the stellar objects used for the experiments.

\noindent
\textbf{\ttfamily galaxies\_output.csv} keeps the predictions generated with the DCMDN in the first experiment.

\noindent
\textbf{\ttfamily quasars\_output.csv} keeps the predictions generated with the DCMDN in the second experiment.

\noindent
\textbf{\ttfamily mixed\_output.csv} keeps the predictions generated with the DCMDN in the third experiment.

\section{SQL-queries}\label{sec:queries}
The following queries have been used to generate the galaxies and stars catalogs via CasJobs:

\begin{minipage}{1.0\columnwidth}
\begin{lstlisting}[
           language=SQL,
           showspaces=false,
           basicstyle=\ttfamily,
           numbers=left,
           numberstyle=\footnotesize,
           commentstyle=\color{gray},
           title="Query used to create the galaxies catalog"
        ]
SELECT TOP 200000
  p.objid,p.ra,p.dec,
  p.u,p.g,p.r,p.i,p.z,
  p.psfMag_u, p.psfMag_g,
  p.psfMag_r, p.psfMag_i,
  p.psfMag_z, p.modelMag_u, 
  p.modelMag_g, p.modelMag_r,
  p.modelMag_i, p.modelMag_z, 
  s.specobjid, s.class, 
  s.z AS redshift 
INTO mydb.DR9_galaxies_with_modMag 
FROM PhotoObj AS p
JOIN SpecObj AS s ON
     s.bestobjid = p.objid
WHERE s.z BETWEEN 0 AND 6.0
  AND s.class = 'GALAXY'
ORDER BY NEWID()
\end{lstlisting}

\begin{lstlisting}[
           language=SQL,
           showspaces=false,
           basicstyle=\ttfamily,
           numbers=left,
           numberstyle=\footnotesize,
           commentstyle=\color{gray},
           title="Query used to extract stellar sources from the SDSS"
        ]
SELECT TOP 200000
   p.objid,p.ra,p.dec,
   p.u,p.g,p.r,p.i,p.z,
   p.psfMag_u, p.psfMag_g,
   p.psfMag_r, p.psfMag_i,
   p.psfMag_z, p.modelMag_u, 
   p.modelMag_g, p.modelMag_r,
   p.modelMag_i, p.modelMag_z, 
   s.specobjid, s.class, 
   s.z AS redshift 
INTO mydb.DR9_stars 
FROM PhotoObj AS p
JOIN SpecObj AS s ON
     s.bestobjid = p.objid
WHERE s.class = 'STAR'
ORDER BY NEWID()
\end{lstlisting}
\end{minipage}

\end{document}